\documentclass[a4paper,twoside,twocolumn,english,showpacs]{revtex4}

\usepackage{times}
\usepackage{amssymb}
\usepackage{graphicx}
\usepackage{amsmath}
\usepackage{color}
\usepackage{ulem}

\newcommand{\note}[1]{}

 \newcommand{\eref}[1]{(\ref{#1})} %short form
\newcommand{\bm}[1]{\boldsymbol{#1}}
\newcommand{\etal}{\textit{et al.}}

\newcommand{\rC}{\text{\bf{C}}} % Classical region
\newcommand{\rI}{\text{\bf{I}}} % Quantum region
\newcommand{\ecut}{\epsilon_{\rm cut}} 
\newcommand{\tecut}{ {\epsilon}_{\rm cut}}

\newcommand{\PC}{\mathcal{P}_{\rC}}

\newcommand{\cf}{\psi_{\rC}} 
\def\x{\mathbf{x}}

\def\k{\mathbf{k}}
\newcommand{\CF}{c-field}

\newcommand{\xa}{(\x)}

\newcommand{\eq}[2]{\begin{equation}\label{#1}#2 \end{equation}}
\newcommand{\eqn}[1]{\begin{eqnarray}#1 \end{eqnarray}}

\newcommand{\kk}{\mathbf{k}}

\begin{document}

\title{Numerical method for evolving the dipolar projected Gross-Pitaevskii equation }

%\author{P. B. Blakie}
%\affiliation{Jack Dodd Centre for Quantum Technology, Department of Physics, University of Otago, New Zealand}
%\author{C. Ticknor}
%\affiliation{ARC Centre of Excellence for Quantum-Atom Optics and Centre for Atom Optics and Ultrafast Spectroscopy,
%Swinburne University of Technology, Hawthorn, Victoria 3122, Australia}
%\author{A.S. Bradley}
%\affiliation{Jack Dodd Centre for Quantum Technology, Department of Physics, University of Otago, New Zealand}
%\author{A. Martin}\affiliation{School of Physics, University of Melbourne, Parkville, VIC 3010, Australia}
%\author{M.J. Davis}
%\affiliation{ARC Centre of Excellence for Quantum-Atom Optics, School of Physical Sciences,  University of Queensland, Brisbane, QLD 4072, Australia}
%\author{Y. Kawaguchi}
%\affiliation{Department of Physics, University of Tokyo, Tokyo 113-0033, Japan}

\author{P. B. Blakie$^1$, C. Ticknor$^2$, A. S. Bradley$^1$, A. M. Martin$^3$, M. J. Davis$^4$, and Y. Kawaguchi$^5$\\ \vspace{6pt} 
$^1$Jack Dodd Centre for Quantum Technology, Department of Physics, University of Otago, PO Box 56, Dunedin New Zealand\\
$^2$ARC Centre of Excellence for Quantum-Atom Optics and Centre for Atom Optics and Ultrafast Spectroscopy,
Swinburne University of Technology, Hawthorn, Victoria 3122, Australia\\
$^3$School of Physics, University of Melbourne, Parkville, VIC 3010, Australia\\
$^4$The University of Queensland, School of Mathematics and Physics, ARC Centre of Excellence for Quantum-Atom Optics, Qld 4072, Australia\\
$^5$Department of Physics, University of Tokyo, Tokyo 113-0033, Japan}

%\affiliation{  Jack Dodd Centre for Quantum Technology, Department of Physics, University of Otago, Dunedin,
%New Zealand}

\date{\today}

\begin{abstract} 
We describe a  method  for evolving the projected Gross-Pitaevskii equation (PGPE) for an interacting Bose gas in a harmonic oscillator potential, with the inclusion of a long-range dipolar interaction. The central difficulty in solving this equation is the requirement that the  field is restricted to a small set of prescribed modes that constitute the low energy \textit{c-field} region of the system.
We present a scheme, using a Hermite-polynomial based spectral representation, that precisely implements this mode restriction and allows an efficient and accurate solution of the dipolar PGPE. We introduce a set of auxiliary oscillator states to perform a Fourier transform necessary to evaluate the dipolar interaction in reciprocal space. We extensively characterize the accuracy of our approach, and derive Ehrenfest equations for the evolution of the angular momentum. 
\end{abstract}

\maketitle 
\section{Introduction}
 The phenomenal recent progress in experimental efforts to produce quantum degenerate dipolar  gases \cite{Stuhler2005a,Sage2005a,Kleinert2007a,Lahaye2007a,Ni2008a} has brought  these systems to the forefront of atomic and condensed matter physics, driven by a broad range of exciting applications
\cite{Baranov2002a,Goral2002a,DeMille2002a,ODell2003a,Kawaguchi2006a,Rabl2006a,Buchler2007a,Ticknor2008b}. 
 Although extensive work has been done on theory for the  $T=0$ dipolar system (e.g. see \cite{Goral2000a,Kawaguchi2006a,Ronen2006b,Kawaguchi2006b,Kawaguchi2007a,Ronen2007a,Ticknor2008a,Wilson2009a,Lahaye2009a,Parker2009a}), a general finite temperature theory has yet to be established. The long-range character of the dipole-dipole interaction has made the development of finite temperature methods more challenging. For example,  meanfield treatments (which have served as the workhorse theory for Bose gases with short-range interactions) have only been applied to the dipolar gas with additional approximations made to the treatment of  exchange interactions  \cite{Ronen2007b}, and quantum Monte Carlo calculations are limited to small numbers of particles \cite{Nho2005a}.

Recently various classical field methods have become popular in the description of ultra-cold Bose gases interacting with short range interactions \cite{Steel1998a,Sinatra2001a,Goral2001a,Davis2001a,Gardiner2003a,Lobo2004a,Polkovnikov2004a}. The appeal of these methods is that the dynamics of the modes are treated non-perturbatively so that non-equilibrium situations or strongly fluctuating equilibrium systems (e.g.~see \cite{Davis2006a}) can be accurately simulated. 
In Ref.~\cite{Blakie2005a} we have developed a quantitative classical field formalism referred to as \textit{c-field theory} \cite{cfieldRev2008}, for which the projected Gross Pitaevskii equation (PGPE) is the underlying equation of motion. This approach has found good agreement with experiment in the critical region of the condensation transition \cite{Davis2006a}, and has seen numerous applications to regimes where traditional meanfield methods are inapplicable (e.g.~see \cite{Simula2006a,Bezett2008a}).
A key component of c-field theory (and the primary distinction from other finite temperature classical field theories  \cite{Goral2001a}) that enables it to be applied to the quantitative description of experiments is the use of a projector, i.e. the explicit restriction of our description to the low energy modes of the system. 
%For typical regimes of interest  of order a thousand modes of the system are sufficiently highly occupied to  be treated using a classical field approach \cite{Blakie2007a}.   

In the literature various numerical techniques have been developed for for solving the ($T=0$) dipolar Gross-Pitaevskii equation, such as Crank-Nicholson \cite{Lahaye2009a}, Fourier pseudospectral \cite{Xiong2009a}, split-operator Fourier transform \cite{Goral2000a} and split-step Fourier transform \cite{Parker2009a,Ronen2006a} methods. 
Underlying all of these approaches is the use of a uniform spacial grid which enables the efficient evaluation of the dipolar term with Fast Fourier transforms.  For accurate simulation of 3D dipolar gases these approaches require $\sim10^6$ spatial grid points.
% However, in most of these realizations hundreds of thousands or millions of points, and hence independent modes of the field, are used. 
In finite temperature applications the number of grid points corresponds to the number of modes that are thermally accessible, and the aforementioned approaches tend to have orders of magnitude too many modes. Indeed, for typical experimental situations of the order of a  few thousand modes are appropriate to be described by the PGPE \cite{Blakie2007a}. In previous work \cite{Blakie2008a} we have found that a practical way to enforce this restriction is by using a numerical approach based on a spectral representation \cite{Dion2003a,Bao2005a}. 
%
%As discussed above, for the PGPE we require careful control over the modes in the simulation and for typical physical situations we wish to work with a few thousand of the lowest energy modes of the system for which a spectral method that precisely implements these modes is most ideal.   

In this paper we develop the numerical underpinnings of a c-field theory  for the dipolar Bose gas by introducing a suitable spectral  technique for solving the dipolar PGPE. 
%This approach includes direct and exchange interactions, beyond meanfield physics (shown to be important in the dipolar system \cite{Astrakharchik2007}), finite temperature effects, and is valid in the critical regime.
The outline of this paper is as follows.
 In Sec.~\ref{SEC:Numerics} we discuss the dipolar PGPE and the spectral representation necessary to implement the explicit  projection.
 In Sec.~\ref{oldPGPE} we briefly review the PGPE algorithm for the trapped Bose gas with contact interactions, before presenting our extension to the dipolar case in Sec.~\ref{dipPGPE}. In Sec.~\ref{accuracy} we present results characterizing the accuracy of our scheme, making comparison to some exactly known matrix elements and other results in the literature. We also examine the convergence of our calculations of  equilibrium properties to provide evidence that the scheme we have developed is suitable to making reliable physical predictions. 
    
\section{Formalism: Dipolar PGPE}\label{SEC:Numerics}
\begin{figure}
\includegraphics[%
  width=3.3in,
  keepaspectratio]{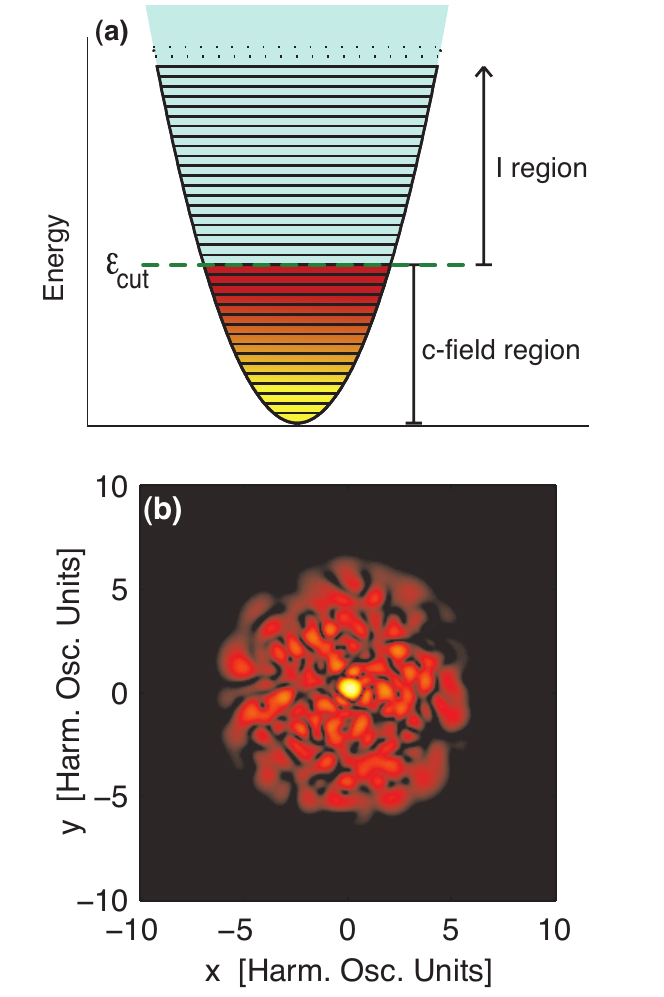}
\caption{\label{Fig:Classicalregion} (a) Schematic diagram showing the c-field ($\rC$) and incoherent ($\rI$) regions of the single particle spectrum for a harmonically
trapped Bose gas. The energy $\epsilon_{{\rm cut}}$ is usually chosen so that the average
number of particles in the modes at the cutoff is   $n_{{\rm cut}}\sim1$.
(b) A typical example of an instantaneous c-field density slice for a dipolar matterwave with  $\ecut=23$.
}
\end{figure}

Our interest is in a system of bosonic particles confined in a harmonic potential, described by the single particle Hamiltonian 
\begin{eqnarray}
H_0&=& 
-\frac{1}{2}\nabla^{2}+V_{0}(\mathbf{x}),\\
V_{0}(\mathbf{x})&=& \frac{1}{2}\sum_{j=1}^3\lambda_j^2x_j^2,
\end{eqnarray} 
where $\lambda_j=\omega_j/\omega$ is the relative trap frequency in each direction $j=\{x,y,z\}$. To obtain this dimensionless form we have used harmonic oscillator units of length $x_0=\sqrt{\hbar/m\omega}$, energy $E_0=\hbar\omega$ and time $t_0=1/\omega$, with $m$ the particle mass and $\omega$ a convenient reference frequency.

Near thermodynamic equilibrium the low energy modes of the system are highly occupied and their dynamics are dominated by classical fluctuations. This observation is at the heart of the c-field technique, and phenomenologically motivates the replacement of the quantum field operator for these modes by a classical field, i.e.~$\hat{\psi}_{\rC}\to\cf$. This replacement can be rigorously justified via a Wigner representation of the many-body density matrix, e.g. see Ref.~\cite{cfieldRev2008}.
However, an immediate consequence of this development is that the c-field formalism must be restricted to the  low energy modes of the system where this field replacement is valid (i.e. the c-field region, $\rC$, shown schematically in Fig.~\ref{Fig:Classicalregion}(a)). To formalize this restriction we introduce a projector, $\PC$\begin{eqnarray}
\PC\{ F(\mathbf{x})\}&\equiv&\sum_{n\in\rC}\phi_{n}(\mathbf{x)}\int d^{3}\mathbf{x}'\phi_{n}^{*}(\mathbf{x'})F(\mathbf{x'}),\label{eq:projector}\\
\rC&=&\{n: \epsilon_{n}\leq \ecut\},\label{eq:Cset}
\end{eqnarray}
where $\phi_{n}(\mathbf{x})$ and $\epsilon_n$ are eigenstates of  $H_{0}$, i.e. 
\begin{equation}
\epsilon_n\phi_n(\x)=H_0\phi_n(\x),
\end{equation}
 and the (single particle) energy cutoff, $\ecut$, is the single parameter we use to define the c-field region \cite{note1}.
The action of $\PC$ in Eq.~(\ref{eq:projector}) is thus to project
the arbitrary function $F(\mathbf{x})$ into the c-field region.

The equation of motion for the c-field treatment of a Bose gas is the projected Gross-Pitaevskii equation (PGPE). For the case of a gas of particles interacting via short range and long range dipole interactions, the PGPE takes the dimensionless form 
\begin{eqnarray}
i\frac{\partial\cf}{\partial t} & = &
H_{0}\cf 
 +\, {\PC}\bigg\{C|\cf(\x)|^{2}\cf(\x) \nonumber \\
 &&+\int d^3x'\,V_{D}(\x-\x')|\cf(\x')|^2\cf(\x)\bigg\},\label{eq:PGPE}\end{eqnarray}
where
\begin{eqnarray} 
V_{D}(\mathbf{x})&=& D \frac{1-3\cos^2\theta}{r^3},
\end{eqnarray} 
is the dipole interaction potential with $r=\sqrt{x^2+y^2+z^2}$ and $\theta$ the angle between $\x$ and the $z$ axis (the axis along which the dipoles are polarized). Here we have introduced the dimensionless $s$-wave (contact) interaction parameter $C=4\pi aN_{\rC}/x_0$, with $a$ the $s$-wave scattering length, and the dimensionless dipole interaction parameter $D=N_{\rC}d^2m/\hbar^2x_0$, with $d$ the dipole moment.
For convenience we take the field $\cf$ to be normalized to unity so that the 
number of c-field atoms, $N_{\rC}$, appears explicitly in the interaction parameters. 

The usual strategy for dealing with the dipolar interaction is to make use of the Fourier transformed density and dipolar interaction potential
\begin{eqnarray}
\tilde{n}(\k) &\equiv&  \int d^3x\,e^{-i\k\cdot\x}|\cf(\x)|^2,\\
\tilde{V}_D(\k) &\equiv&  \int d^3x\,e^{-i\k\cdot\x}V_D(\x),\\
&=& -\frac{4\pi D}{3}\left[1-3\cos^2\theta_k\right],
\end{eqnarray}
where $\theta_k$ is the angle between $\mathbf{k}$ and the $k_z$ axis. Thus making use of the convolution theorem we have 
\begin{eqnarray}
\Phi\xa &\equiv&\int d^3x'\,V_{D}(\x-\x')|\cf(\x')|^2,\label{Phi}\\
&=& \int d^3k\,e^{i\k\cdot\x}\tilde{V}_D(\k)\tilde{n}(\k)\label{Phi2}.
\end{eqnarray}
The main concern of this paper is to develop a suitable method for evaluating $\Phi(\x)$ in a manner appropriate for use in the PGPE formalism.  We emphasize that the modes of the system are of central importance in the PGPE and care must be taken in numerical implementations to ensure the modes are faithfully represented. This point is made clear with reference to Fig.~\ref{Fig:Classicalregion}(b), which shows a snapshot of the c-field density and reveals the appreciable occupation of every mode in the c-field region.

We also note that the energy functional for the dipolar PGPE is 
 \begin{eqnarray}
E[\cf] & = &
\int d^3x\,\cf^*H_{0}\cf+ \frac{1}{2}\int d^3x\, C|\cf(\x)|^{4}\label{eq:EPGPE} \\
 &&+\frac{1}{2}\int d^3x\,\Phi(\x)|\cf(\x)|^2,\nonumber 
 \end{eqnarray}
 which forms an important constant of motion for the system. In a similar manner to how we dealt with the dipolar part of the PGPE, it is convenient to evaluate the dipolar energy term in Fourier space as 
% \footnote{We note that our result $\frac{1}{2}\int d^3k\, \tilde{V}_D(\k)\tilde{n}(\k)\tilde{n}(-\k)$ differs from that given in \cite{Ronen2006a} where $\tilde{n}(\k)^2 $ appears in the integrand. For the case of symmetric states considered in that paper the two forms are equivalent.}
 \begin{equation}
 \frac{1}{2}\int d^3x\,\Phi(\x)|\cf(\x)|^2=\frac{1}{2}\int d^3k\, \tilde{V}_D(\k)\tilde{n}(\k)\tilde{n}(-\k). 
 \end{equation}

\subsection{Spectral representation}\label{sspecrep}
It is most convenient to expand the \CF\ in a spectral basis of the single particle states, i.e.
\begin{equation}
  {\cf}( {\mathbf{x}},  t)=\sum_{n\in\rC}c_{n}(  t)\, \phi_{n}( {\mathbf{x}}),\label{eq:psibasis}\end{equation}
 where the $\{c_n\}$ are complex amplitudes. 
 The projection
is explicitly implemented by limiting the summation indices in \eref{eq:psibasis} to the set of values specified in Eq.~(\ref{eq:Cset})
defining the c-field region.

\subsection{Mode evolution}\label{smodeevol}
Having used the modes of $ {H}_0$ as the spectral basis and to realize the projector,  we follow the Galerkin approach  (i.e.~projecting   Eq.~\eref{eq:PGPE} on to our spectral basis) to obtain the evolution equation for the mode amplitudes
\begin{eqnarray}\label{eq:GPEshobasis}
\frac{\partial c_{n}}{\partial  t} & = & -i\left[ \epsilon_{n}c_n +G_{n}\right],\end{eqnarray}
 where
\begin{eqnarray}\label{eq:GNL}
G_{n}&\equiv&\int d^{3} {\mathbf{x}}\: \phi_{n}^{*}( {\mathbf{x}})\left[C|  {\cf}|^{2} +\Phi\right] {\cf}( {\mathbf{x}}, {t}),\end{eqnarray}
is the nonlinear matrix element. Once these nonlinear matrix elements are evaluated, the evolution of the system can be calculated using  numerical algorithms for  systems of ordinary differential equations, e.g.~the Runge-Kutta algorithm. Since this is a well-understood area of numerical mathematics we do not concern ourselves with the details of the propagation algorithm, but instead focus on evaluating Eq.~\eref{eq:GNL}.  

In principle the nonlinear matrix elements between spectral basis functions can be computed exactly. Defining 
%\begin{equation}
%I_{npqr}\equiv \int d^3x\,\phi_n^*(\x)\left[C\phi_p^*(\x)+\int d^3x'\,V_D(\x-\x')\phi_p^*(\x')\phi_q(\x')\right]\phi_r^*(\x).
%\end{equation}
\begin{eqnarray}
I_{npqr}&\equiv& \int d^3x\,\phi_n^*(\x)\left[C\phi_p^*(\x)\phi_q(\x)\phi_r(\x) + \right. \\
&&\int d^3x'\,\left.V_D(\x-\x')\phi_p^*(\x')\phi_q(\x')\phi_r(\x)\right],\nonumber
\end{eqnarray}
which can be calculated analytically (c.f Appendix \ref{dipoleME}), 
and expanding the c-field in terms of its spectral representation we see that
\begin{equation}
G_n=\sum_{\{p,q,r\}\in\rC}I_{npqr}c_p^*c_qc_r.
\end{equation}
While being exact, evaluating this expression is prohibitively slow, requiring $O(M^4)$ operations, where $M$ is the number of modes in the c-field region. In contrast, the approach we develop here is $O(M^{4/3})$, and thus suitable for simulating real systems in a reasonable amount of time (e.g. simulations of the order of hours to days on a commodity PC). 

%We can point out the central issue for numerical implementation. Expanding the fields in expression \eref{eq:GNL} into the mode basis we obtain
%\begin{equation}
%G_{n}=\sum_{pqr}\left\{\int d^{3} {\mathbf{x}}\:  \phi_{n}^{*}( {\mathbf{x}}) \phi_{p}^{*}( {\mathbf{x}}) \phi_{q}^{}( {\mathbf{x}}) \phi_{r}^{}( {\mathbf{x}})\right\}\: c^*_pc_qc_r.
%\end{equation}
%While the matrix elements within the brackets can be exactly calculated in advance,  computing all $G_n$ values  using this expression requires $O(M^4)$ floating point operations, where $M$ is the number of \CF\ region modes. Such scaling would be prohibitive for performing realistic calculations. In what follows we show how to compute these matrix elements with a scheme that only requires $O(M^{4/3})$ operations. Such spectral representations have also been considered for the zero-temperature (non-projected) Gross-Pitaevskii equation in references \cite{Dion2003a,Bao2005a}.

\subsection{Separability} \label{sAsep}
In what follows we take the trap to be isotropic, and set all $\lambda_j=1$, for simplicity of notation \cite{note2}.
 An important feature of the basis states (i.e.~eigenstates of $ {H}_0$) is that they are separable into 1D eigenstates, i.e.
\begin{eqnarray}
 {\phi}_n( {\mathbf{x}})&\leftrightarrow& \varphi_{\alpha}( {x}) \varphi_{\beta}( {y}) \varphi_{\gamma}( {z}),\label{EQ:PW1Dstates}\\
 {\epsilon}_n&\leftrightarrow& {\varepsilon}_{\alpha}+ {\varepsilon}_{\beta}+ {\varepsilon}_{\gamma},\label{EQ:PW1Denergies}\\
c_n&\leftrightarrow&c_{\alpha\beta\gamma}, \label{EQ:PW1Damps}
\end{eqnarray} 
where $\{ \varphi_{\alpha}(  x)\}$ are eigenstates of the 1D harmonic
oscillator Hamiltonian, i.e.\begin{equation}
\left[-\frac{1}{2}\frac{d^{2}}{d  x^{2}}+\frac{1}{2} x^{2}\right] \varphi_{\alpha}(  x)= \varepsilon_{\alpha} \varphi_{\alpha}(  x),\label{eq:sho1D}\end{equation}
 with eigenvalue $ \varepsilon_{\alpha}=(\alpha+\frac{1}{2})$, for $\alpha$ a non-negative integer.

For clarity we use greek subscripts to label the 1D eigenstates, so that the specification of the \CF\  region in \eref{eq:Cset} becomes \begin{equation}
\rC=\{\alpha,\beta,\gamma: \varepsilon_{\alpha}+ \varepsilon_{\beta}+ \varepsilon_{\gamma}\le\tecut\}.\label{sepCR}
\end{equation} 
Within the \CF\ region there exists $M_x$ ($\approx {\epsilon}_{\rm{cut}}$) distinct 1D eigenstates (i.e.~$ {\varphi}_{\alpha}$) in each direction, and thus 
\begin{equation}M\approx \frac{1}{6}M_x^3,\label{eqM}
\end{equation}
 3D basis states ($ {\phi}_n$) in the \CF\ region.

\section{Review of standard PGPE algorithm} \label{oldPGPE} 
We first begin by reviewing the PGPE algorithm we have developed for the case of local interactions. This algorithm uses Gauss-Hermite quadrature  to evaluate the  (local) nonlinear term exactly in an efficient manner. For a complete account we refer the reader to Ref.~\cite{Blakie2008a}.

\subsection{Evaluating the matrix elements} \label{sec:matrixEval}
To begin we note the harmonic oscillator
states are of the form 
\begin{equation}
\varphi_{\alpha}(  x)=h_{\alpha} H_{\alpha}(  x)e^{-x^{2}/2},
\end{equation}
where $h_{\alpha}=[2^\alpha\alpha!\sqrt{\pi}]^{-1/2}$ is a normalization constant, and $H_{\alpha}( x)$ is a Hermite polynomial of degree $\alpha$, defined by the recurrence relation
\begin{equation}
H_{\alpha+1}(x)=2xH_{\alpha}(x)-2\alpha H_{\alpha-1}(x),\quad \alpha=1,2,\ldots,
\end{equation}
with $H_0(x)=1$ and $H_1(x)=2x$.

Thus, the field (at any instant of time) can be written as\begin{equation}
  \cf( {\mathbf{x}},  t)=Q(  x,  y,  z)e^{-(  x^{2}+  y^{2}+  z^{2})/2},\label{eq:PsiQpoly}\end{equation}
 where 
 \begin{equation}Q(  x,  y,  z)\equiv \sum_{\{\alpha\beta\gamma\}\,\in\,\rC}c_{\alpha\beta\gamma}(  t)\,h_{\alpha}H_{\alpha}(  x)h_{\beta}H_{\beta}(  y)h_{\gamma}H_{\gamma}(  z),
 \end{equation}is a polynomial that, as a result of the cutoff,
is of maximum degree $M_x-1$ in the independent variables.

Similarly, it follows that because the interaction term (\ref{eq:GNL}) is fourth
order in the field, it can be written in the form\begin{equation}
G_{\alpha\beta\gamma}=\int d^{3} {{x}}\: e^{-2(  x^{2}+  y^{2}+  z^{2})}P_{\alpha\beta\gamma}(  x,  y,  z),\label{eq:FNLquad}\end{equation}
 where 
\begin{eqnarray}P_{\alpha\beta\gamma}(  x,  y,  z)&\equiv& C\,h_{\alpha}H_{\alpha}(  x)h_{\beta}H_{\beta}(  y)h_{\gamma}H_{\gamma}(  z)\nonumber\\ &&\times|Q(  x,  y,  z)|^2Q(   x,  y,  z),\label{Ppoly}
\end{eqnarray}
 is a polynomial of maximum degree $4\,(M_x-1)$
in the independent variables. To evaluate these integrals, we note the general form of the $N_Q$ point Gauss-Hermite quadrature
\begin{equation}
\int_{-\infty}^{+\infty}d {x}\,w( {x})f( {x})\approx\sum_{j=1}^{N_Q}w_jf( {x}_j),
\end{equation}
where $w( {x})$ is a Gaussian weight function, and the $N_Q$ values of $w_j$ and $x_j$ are the quadrature weights and roots, respectively. This quadrature   is exact if $f( {x})$ is a polynomial of maximum degree $2N_Q-1$.

 Identifying the exponential term in \eref{eq:FNLquad} as
the  weight function for quadrature,  the integral
can be exactly evaluated using a three-dimensional spatial 
grid of $8\,(M_x-1)^3$ points (i.e. $2\,(M_x-1)$ points in each direction \cite{note3}), i.e.
\begin{equation}
G_{\alpha\beta\gamma} =\sum_{ijk}w_iw_jw_kP_{\alpha\beta\gamma}( {x}_i, {x}_j, {x}_k),\label{Gabc}
\end{equation}
where $  x_i$ and $w_i$ are the $2\,(M_x-1)$ roots and weights of the 1D Gauss-Hermite quadrature with weight function $w(  x)=\exp(-2  x^2)$. % \cite{abramowitz1964a}.
 Note, that the isotropy of the trapping potential (for the numerical examples considered in this paper) results in identical quadrature grids in all spatial directions in our example.

\subsection{Overview of the numerical algorithm}\label{SEC:HARMnummeth}
Here we briefly overview how the quadrature described  above can be efficiently implemented numerically. We require the transformation matrices, given by
 1D basis states evaluated on the quadrature grid, i.e.
\begin{equation}
U_{i\alpha}= \varphi_{\alpha}( {x}_i),\label{harmU}
\end{equation}
to be pre-calculated. Because the transformations are block diagonal, i.e.~applied across the directions independently at computational cost $O(M_x^4)=O(M^{4/3})$ (see Eq.~(\ref{eqM})), we will make use of the simplifying notation\begin{equation}
\sum_{\{\alpha\beta\gamma\}\,\in\,\rC}U_{i\alpha}U_{j\beta}U_{k\gamma}\,c_{\alpha\beta\gamma}(  t)\to \sum_{\bm{\sigma}} { U}_{\bm{s}\bm{\sigma}}c_{\bm{\sigma}},
\end{equation}
where $\bm{\sigma}=\{\alpha\beta\gamma\}$ and $\bm{s}=\{ijk\}$, and it is understood
that  $c_{\bm{\sigma}}=c_{\alpha\beta\gamma}$, and $U_{\bm{s}\bm{\sigma}}=
U_{i\alpha}U_{j\beta}U_{k\gamma}$.

 Starting from the basis set representation of the field (i.e. $\{c_{\alpha\beta\gamma}\}$) at an instant of time $  t$, the steps for calculating the matrix elements are as follows:
\begin{enumerate}
\item[Step 1:] Transform from spectral to spatial representation:
\begin{eqnarray}
\cf(\x_{\bm{s}})&=&\sum_{\bm{\sigma}} U_{\bm{s}\bm{\sigma}}c_{\bm{\sigma}},
\end{eqnarray}
where $ {\mathbf{x}}_{\bm{s}}=( {x}_i, {x}_j, {x}_k)$.

\item[Step 2:]  The quadrature integrand of the nonlinear matrix element  \eref{eq:GNL}  is constructed by appropriately dividing by the weight function and pre-multiplying by the weights \cite{note4}, i.e.
\begin{eqnarray} 
g(\x_{\bm{s}})&=&w_{\bm{s}}e^{2|\x_{\bm{s}}|^2}C|  {\cf}(\x_{\bm{s}}, {t})|^{2}  {\cf}(\x_{\bm{s}}, {t}),
\end{eqnarray}
where $w_{\bm{s}}=w_iw_jw_k$.

\item[Step 3:]  The inverse transform of $g(\x_{\bm{s}})$ yields the desired matrix elements:
\begin{eqnarray}   
G_{\bm{\sigma}} &=& \sum_{\bm{s}}U_{\bm{s}\bm{\sigma}}^*g(\x_{\bm{s}}).
\end{eqnarray}
\end{enumerate}
  The slowest step in this procedure is carrying out the basis transformation (steps 1 and 3), which requires  $O(M^{4/3})$ floating point operations when carried out as a series of matrix multiplications.  Thus, the overall algorithm is $O(M^{4/3})$.
  
\section{Extension to calculate the dipolar term}\label{dipPGPE}
To treat the dipolar term we need to augment step 2 in the standard harmonic PGPE algorithm (see Sec.~\ref{SEC:HARMnummeth}). To do this we want to Fourier transform the density associated with ${\cf}( {\x}_{\bm{s}}, {t})$ to form Eq.~(\ref{Phi2}). It is not convenient to use a fast Fourier transform because ${\cf}( {\x}_{\bm{s}}, {t})$ is evaluated on a nonuniform grid (i.e.~quadrature grid). Interpolation to a uniform grid would be computationally expensive and would introduce a source of considerable error, especially since the quadrature grids tend to be quite sparse (see discussion in Sec.~\ref{secFFT}).

Here we show how an auxiliary harmonic oscillator basis can be used to perform the Fourier transform exactly. Following similar arguments to those made in Sec.~\ref{sec:matrixEval}, the \CF\ density, $n\xa=|\cf\xa|^2$,  is of the form
\begin{equation}
n\xa=R(  x,  y,  z)e^{-(  x^{2}+  y^{2}+  z^{2})},\label{eq:PsiRpoly}
\end{equation}
where $R$ is a polynomial of maximum degree $2(M_x-1)$ in the independent variables.

Introducing a set of \textit{auxiliary} harmonic oscillator states,
\begin{equation}
{\chi}_\alpha(x) = \bar{h}_\alpha\bar{H}_\alpha(x)e^{-x^2},
\end{equation}
which differ from the spectral basis oscillator states by a factor of 2 in the argument of the exponential (chosen to match the exponential part of   Eq.~(\ref{eq:PsiRpoly})). Indeed, these states are eigenstates of the operator
\begin{equation}
\bar{H}_x=\left[-\frac{1}{2}\frac{d^2}{dx^2}+2x^2\right],
\end{equation}
i.e.~harmonic oscillator with twice-as-tight trapping potential, and expressions for $\bar{h}_\alpha$ and $\bar{H}_\alpha(x)$ can be obtained by noting that these modes relate to the usual dimensionless oscillators by a simple scaling $ {\chi}_\alpha(x) =2^{1/4}\varphi_{\alpha}(\sqrt{2}x)$.

The auxiliary oscillator states form an orthonormal basis, and because of their appropriate exponential factor, we can exactly represent the density (\ref{eq:PsiRpoly}) as
\begin{eqnarray} n({\bf x})&=&\sum_{\bm{\sigma}}d_{\bm{\sigma}}\chi_{\bm{\sigma}}({\bf x})
\end{eqnarray}
where $d_{\bm{\sigma}}\leftrightarrow d_{\alpha\beta\gamma}$ is a set of $8M_x^3$ real coefficients,  with $\chi_{\bm{\sigma}}(\x)=\chi_\alpha(x)\chi_\beta(y)\chi_\gamma(z)$. Indeed, because the $\{\chi_{\bm{\sigma}}\}$ are an orthonormal basis, we have that
\begin{eqnarray} 
d_{\bf{\sigma}}&=&\int d^3x\,\chi^*_{\bm{\sigma}}({\bf x})n({\bf x}),\label{dsig}
\\  
&=&\int d^3x\,e^{-2{|\x|}^2}S_{\sigma}({\bf x}),\label{dsig2}
\end{eqnarray}
where in the second line we have collected exponential and polynomial terms separately, with 
\begin{eqnarray} 
S_{\sigma}\xa=e^{2{|\x|}^2}\chi_\sigma^*({\x})n\xa,\label{sig}
\end{eqnarray}
 a polynomial of degree $4(M_x-1)$ in the independent variables. Thus the integration 
 (\ref{dsig2}), like that in Eq.~(\ref{eq:FNLquad}), has same weight function and maximum degree of polynomial order. Thus Eq.~ (\ref{dsig2}) can be calculated exactly with the same quadrature  (i.e.~roots $\{x_i\}$ and weights $\{w_i\}$) as used in Eq.~(\ref{Gabc}), i.e.
\begin{eqnarray} 
d_{\bm{\sigma}}&=&\sum_{\bm{s}}w_{\bm{s}}\,S_{\bm{\sigma}}(x_s).
\end{eqnarray}

 The harmonic oscillator states are eigenstates of the Fourier transform operator with eigenvalue $(-i)^{\alpha}$, i.e. 
\begin{equation}
 {\chi}_{\alpha}{(k_x)}=(-i)^{-\alpha}\frac{1}{(2\pi)^{1/2}}\int d{x}\,e^{-ik_x {x}}{\chi}_{\alpha}({x}).
 \end{equation}
 Thus knowledge of the basis amplitudes $d_{\alpha\beta\gamma}$ allows us to efficiently and precisely construct the Fourier transform of the classical field density, i.e.
 \begin{eqnarray} 
\tilde{n}(\k)&=&{\sum_{\bm{\sigma}}}d_{\sigma}\,(-i)^{-|\!|\bm{\sigma}|\!|_1}{\chi}_{\bm{\sigma}}({\k}),
 \end{eqnarray}
where $|\!|\bm{\sigma}|\!|_1=\alpha+\beta+\gamma$ is the one norm of $\bm{\sigma}$, noting that $\{\alpha,\beta,\gamma\}$ are non-negative.
We can now construct the integrand of the dipolar interaction term in Fourier space, i.e.~$\tilde{V}_D(\k)\tilde{n}(\k)$ appearing in Eq.~(\ref{Phi2}), which needs to be inverse Fourier transformed to obtain $\Phi\xa$. This can be done using the inverse of the procedure we used to obtain  $\tilde{n}(\k)$, i.e. via the expansion of $\Phi(\x)$ in the auxiliary oscillator states
\begin{eqnarray}
\Phi\xa&\approx&\sum_{\bm{\sigma}} f_{\bm{\sigma}}{\chi}_{\bm{\sigma}}({\x}),\label{Phi3}
\end{eqnarray}
where
\begin{eqnarray}
f_{\bm{\sigma}}=\int d^3k\,(i)^{-|\!|\bm{\sigma}|\!|_1}{\chi}_{\bm{\sigma}}^*({\bf k})\tilde{V}_D(\k)\tilde{n}(\k).\label{fabc}
\end{eqnarray}
%are the amplitudes of $\tilde{V}_D(\k)\tilde{n}(\k)$  represented in the auxiliary harmonic oscillator basis. 
Expression (\ref{Phi3}) is approximate because $\tilde{V}_D(\k)$ is not of the form of a finite-degree polynomial, and thus $\tilde{V}_D(\k)\tilde{n}(\k)$ cannot be represented exactly in the oscillator basis -- an approximation we investigate in Sec.~\ref{accuracy}.

To numerically evaluate the $f_{\alpha\beta\gamma}$ we again make use of a Hermite-Gauss quadrature with roots $\{k_i\}$ and weights $\{\bar{w}_i\}$, i.e.
\begin{eqnarray}
f_{\sigma}&=&\sum_{\bm{s}}\bar{w}_{\bm{s}}T_{\bm{\sigma}}(\k_{\bm{s}}),
\end{eqnarray}
where
\begin{eqnarray}
T_{\bm{\sigma}}(\k)&=&e^{2|\k|^2}{\chi}_{\bm{\sigma}}^*(\k)\tilde{V}_D(\k)\tilde{n}(\k).
\end{eqnarray}
Note the number of $k$-grid quadrature points is in principle arbitrary, but should be at least $2M_x$ in each direction. We can use the number of points to control the accuracy of the matrix element.

\subsection{Spectral dipolar algorithm summary}\label{dipalg}
\begin{itemize}
\item[Step 1:] Transform from spectral to spatial representation:
\begin{eqnarray} 
{\cf}( {\mathbf{x}}_{\bm{s}})&=&\sum_{\bm{\sigma}}U_{\bm{s}\bm{\sigma}}c_{\bm{\sigma}}.
\end{eqnarray} 

\item[Step 2a:] The weighted position density is constructed
\begin{eqnarray} 
f(\x_{\bm{s}})&\equiv& w_{\bm{s}}e^{2|\x_{\bm{s}}|^2}|\cf(\x_{\bm{s}})|^2.
\end{eqnarray}
\item[Step 2b:] We compute the Fourier transformed density as
\begin{eqnarray} 
\tilde{n}(\k_{\bm{t}})&=&\sum_{\bm{s}}W_{\bm{s}\bm{t}}f(\x_{\bm{s}}),
\end{eqnarray}
where $\bm{t}=\{uvw\}$ are the indices which label the Fourier space grid points. Here we have introduced 
the pre-computed  transformation matrix,
\begin{eqnarray}
W_{ir} = \sum_\alpha(-i)^\alpha{\chi}_\alpha(k_r){\chi}_\alpha(x_i).
\end{eqnarray}
 which combines both steps of the Fourier transform into one (i.e.~$n\xa\to d_{\alpha\beta\gamma}$ and $d_{\alpha\beta\gamma}\to \tilde{n}(\k)$).  
% It is important to remark that this basis functions are also their own Fourier transform apart from a prefactor.
\item[Step 2c:] The product with the dipole interaction potential is then formed in Fourier space
\begin{eqnarray}
\tilde{f}(\k_{\bm{t}})&\equiv& \bar{w}_{\bm{t}}e^{2|\k_{\bm{t}}|^2}\tilde{V}_D(\k_{\bm{t}})\tilde{n}(\k_{\bm{t}}).
\end{eqnarray} 
[Or with the replacement $\tilde{V}_D(\k_{\bm{t}})\to\tilde{V}_D^R(\k_{\bm{t}})$, a corrected dipolar interaction, as discussed in Sec.~\ref{MIT}].
\item[Step 2d:] Inverse transforming yields
\begin{eqnarray} 
\Phi(\x_{\bm{s}})&=&\sum_{\bm{t}}W^*_{\bm{s}\bm{t}}\tilde{f}(\k_{\bm{t}}).
\end{eqnarray}
\item[Step 2e:] Short range and dipolar interaction terms are then combined into a single integrand
\begin{eqnarray}  
g(\x_{\bm{s}})&\equiv&w_{\bm{s}}e^{2| \x_{\bm{s}}|^2}  \left[C| {\cf}( \x_{\bm{s}})|^{2}+\Phi(\x_{\bm{s}})\right]{\cf}( \x_{\bm{s}}).
\end{eqnarray}
\item[Step 3:] Inverse transforming this integrand yields the desired matrix elements:
\begin{eqnarray}  
G_{\bm{\sigma}} &=& \sum_{\bm{s}}U_{\bm{s}\bm{\sigma}}^*g(\x_{\bm{s}}).
\end{eqnarray}
\end{itemize}
Steps 1, 2b, 2d, and 3 are $O(M^{4/3})$. Since the algorithm involves twice as many transformations as the non-dipolar PGPE case, each evaluation of the $G_{\bm{\sigma}}$ (and hence each time step) takes approximately twice as long.

\subsection{Possibility of using fast Fourier transformations}\label{secFFT}
Having presented our spectral algorithm we are now able to comment on the alternative procedure of computing $\Phi$ using fast Fourier transformations (FFTs).
To do this requires several modifications to the algorithm, which we briefly summarize. In step one, in addition to computing $\cf(\x_{\bm{s}})$ on the quadrature grid for the short range interaction, we will need a new transformation $\bar{U}_{\bm{s}}$ to obtain $\cf(\bar{\x}_{\bm{s}})$ on the uniformly spaced grid $\{\bar{\x}_{\bm{s}}\}$. Following standard procedures (e.g.~see \cite{Goral2000a}) we can then obtain $\Phi(\bar{\x}_{\bm{s}})$ using two FFTs. This step is more efficient than our procedure using $W_{\bm{s}\bm{t}}$  in the spectral algorithm, but we will likely require more $\bar{\x}_{\bm{s}}$ grid points for the Fourier representation to provide an adequate  representation of trapped field. Additionally, the efficiency of the FFTs is offset by the need to interpolate $\bar{\x}_{\bm{s}}$ back onto a quadrature grid for step 3 (if performed on a uniform grid this last step is highly inaccurate without a prohibitively large number of points). Due to the added complexity of the FFT algorithm, and that approximations occur in the algorithm at several places, we have decided not to investigate this any further in this work.

\section{Accuracy of approach}\label{accuracy}
Step 2d of our numerical algorithm for the dipolar PGPE is approximate and requires investigation to justify that it is sufficiently accurate to be useful.  
The PGPE formalism places strong constraints on the underlying numerical algorithm which restrict how we might improve the accuracy. In particular, the c-field region is defined by $\ecut$, and hence $M_x$ is dictated by the physical system under consideration (i.e. temperature, number of atoms) and is not a parameter that can be arbitrarily varied. 
%So that checking the convergence of results by increasing the size of the spectral representation, $M_x$, is not relevant to the PGPE. 
Instead, for fixed $M_x$, we would like to understand: (i) The accuracy of the matrix elements $G_{\bm{\sigma}}$. (ii) What ways we have for controlling this accuracy? (iii) What level of accuracy is needed for making reliable physical predictions?

Here we investigate two methods of improving the accuracy of the matrix elements. The first method, which we discuss in Sec.~\ref{FQG}, is by increasing the order of the $k$-space quadrature. The second method is to use a modified (finite range) interaction potential, which we present in Sec.~\ref{MIT}. We then characterize the effect of these adjustments using various tests. We finally turn to addressing what level of accuracy is required to make useful predictions with the PGPE theory.

\subsection{Fourier quadrature grid}\label{FQG}

The two quadrature grids $\{\mathbf{x}_{\mathbf{s}}\}$ and $\{\mathbf{k}_{\mathbf{t}}\}$ are central to the computation of the nonlinear matrix elements in our algorithm. Since the weight functions are known for each quadrature they are completely specified by the number of points, i.e.~the parameters
\begin{itemize}
\item[$N_x$:] The number of quadrature points along each direction in the position space $x$ grid. 
%\item[$M_A$: ] The maximum order of auxiliary oscillator state.
\item[$N_k$:] The number of quadrature points along each direction in the Fourier space $k$ grid. 
\end{itemize}
First, we note that for given $\ecut$ (i.e. $M_x$) the transform to $k$-space is exactly invertible (i.e.~$n(\x)\stackrel{W}{\longrightarrow}\tilde{n}(\k)\stackrel{W^\dagger}{\longrightarrow}n(\x)$) if we choose $N_x\ge N_x^0$, $N_k\ge N_k^0$, where we have defined the reference values
\begin{eqnarray}
N_x^0&\equiv&2M_x-1,\\ 
N_k^0&\equiv&2M_x.
\end{eqnarray}
Note, that the invertible requirement is met with $N_k^0\equiv2M_x-1$, but we choose $2M_x$ to avoid having an odd number of points which ensures that there is no quadrature point  at $\k=\bm{0}$ where $\tilde{V}_D$ is singular.
 
With the inclusion of the dipolar potential it is beneficial to increase the number of momentum grid points beyond $N_k^0$ to obtain better accuracy for step 2d. In the results we present below we will indicate the increase in momentum grid points over the reference value as $\Delta N_k$, i.e.
 \begin{equation}
 N_k=N_k^0+\Delta N_k.
 \end{equation}
 We do not alter $N_x$ from the reference value $N_x^0$, as this has no effect on the accuracy of the algorithm.
% We note that $N_k^0$ should be $2M_x-1$ to exactly transform the density to Fourier space and back in the absence of the dipole potential. However preempting our consideration of the dipole potential, we note that odd quadrature grids have a point at the origin where the dipole potential is singular. As a result, odd grids always produce results significantly less accurate than the nearest even grids and for this reason we have taken $N_k^0=2M_x$ will only consider even $\Delta N_k$ here. 
 
\subsection{Corrected dipolar interaction}\label{MIT}
 Ronen \etal\ \cite{Ronen2006a} have demonstrated a useful procedure for improving the convergence of the numerical evaluation of the dipolar term for low energy states   in Bogoliubov calculations. They noted that the poor convergence of this term arises because the Fourier transformed interaction, $\tilde{V}_D(\k)$, is singular at the origin (where $\tilde{n}(\k)$ is typically large) due to the long range character of the interaction. Ronen \etal~suggested the use of the Fourier transform of the dipolar interaction restricted to a spherical domain of size $R$, i.e.~the Fourier transform of
 \begin{eqnarray}
 V_D^R(\x)=\left\{\begin{array}{c c}
               D ({1-3\cos^2\theta})/{r^3} \quad& r<R,   \\
               & \\
               0, & \rm{otherwise}.
          \end{array}\right.
 \end{eqnarray}
 This has the analytic transform
 \begin{equation}
  \tilde{V}_D^R(\k)=\frac{4\pi D}{3}\left(1+3\frac{\cos(Rk)}{R^2k^2}-3\frac{\sin(Rk)}{R^3k^3}\right)(3\cos^2\theta_k-1),
 \end{equation}
 which we shall refer to as the \textit{corrected dipolar interaction}, having the feature that it is less rapidly varying near $\k=\bm{0}$. This approach seems reasonable as we are studying a  trapped system of finite spatial extent, and thus the sharp behavior of the uncorrected potential ($\tilde{V}_D$) at $\k=\mathbf{0}$, arising from  interactions over long length scales, cannot be physically relevant.
 Ronen \etal~justify using $V_D^R$ as it prevents the ``long range interactions between copies of condensates" arising from the periodicity of their Fourier based calculations. 

% A more general explanation is that $V_D^R$ helps improve the convergence by preventing the misrepresentation of high order matrix elements, i.e.~outside those represented by the quadrature,  
More generally, the use of  $V_D^R$ can be justified by noting that sharp features in the interaction potential are not accurately calculated on a finite quadrature grid (or Fourier grid). In practice if these sharp features are left in the numerical calculations they are misrepresented by the finite quadrature and interfere with lower order matrix elements (often referred to as aliasing in the Fourier case), leading to their slow convergence as the number of quadrature points is increased.

% 
%An analytic form of the corrected dipolar interaction is not known for the case of different restriction lengths in each spatial direction, although Ronen \etal~also give a result for an interaction restricted in only one direction.

\subsubsection*{Choice for $R$}
An immediate issue to investigate is the optimal choice of the length scale  $R$.  For the our trapped system the characteristic size is given by the classical turning point $l_{tp}\approx\sqrt{2M_x}$ 
(in computational units), since $\ecut\approx M_x$.

To investigate the accuracy of our algorithm as we vary $R$ used in the corrected dipolar interaction we consider the \textit{pure dipolar matrix element} (for $D=1$):
\begin{equation}
Z_{\bm{\nu}}^{ \bm{\tau}}\equiv\int d^3x\,d^3x'\,\phi^*_{\bm{\tau}}(\x)V_D(\x-\x')|\phi_{\bm{\nu}}(\x')|^2\phi_{\bm{\nu}}(\x).\label{puredipmeZ}
\end{equation}
to be distinguished from the general matrix element which requires four distinct oscillator state labels.
In practice we evaluate this as follows: we take $c_{\bm{\sigma}}=\delta_{\bm{\sigma},\bm{\nu}}$, and then compute the nonlinear matrix elements $G_{\bm{\tau}}$ using our algorithm (see Sec.~\ref{dipalg}), and identify $Z_{\bm{\nu}}^{ \bm{\tau}}=G_{\bm{\tau}}$. These pure dipole matrix elements are useful for characterizing the accuracy of the algorithm and we will make use of these in several applications. It is convenient to indicate the matrix elements under consideration using the notation $Z_{\alpha\beta\gamma}^{\delta\epsilon\zeta}$, as established earlier (E.g.~see Eq.~(\ref{EQ:PW1Dstates})). 

\begin{figure}[htbp] %  figure placement: here, top, bottom, or page
   \centering
   \includegraphics[width=3.2in]{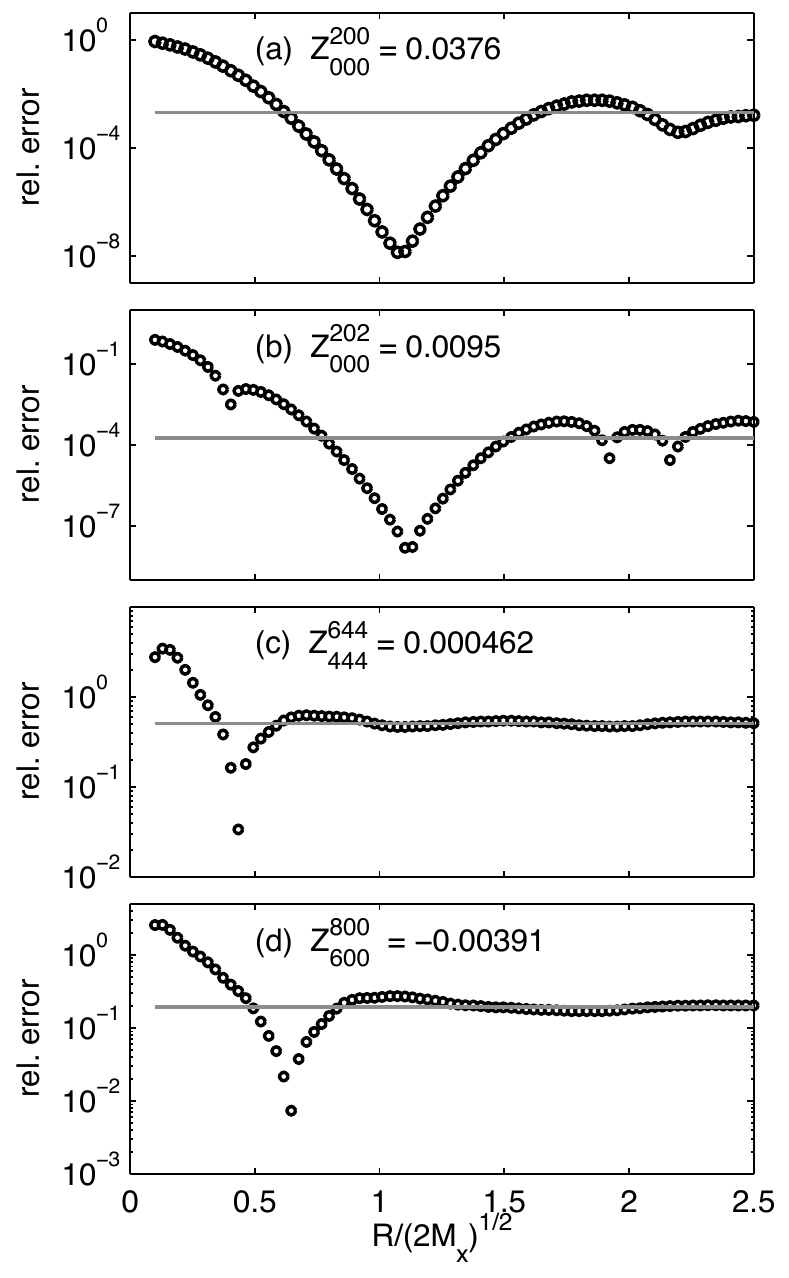} 
   \caption{Relative error in the pure dipole matrix elements (a) $Z^{200}_{000}$, (b)   $Z^{202}_{000}$, (c) $Z^{644}_{444}$ and (d) $Z^{800}_{600}$, as $R$ is varied. Results: (solid line) calculated using  $\tilde{V}_D(\k)$ ; (circles) calculated using  $\tilde{V}_D^R(\k)$. Other parameters are at the reference values with $M_x=16$. }
   \label{fig:R}
\end{figure}

For the purposes of studying the dependence on $R$ we will consider four non-trivial matrix elements shown in Table \ref{tab:ZRresults} for which we calculate the values exactly using an analytic approach discussed in Appendix \ref{dipoleME}. The results of our algorithm are shown in  Fig.~\ref{fig:R} and confirm that the corrected interaction potential, $\tilde{V}_D^R$, has considerable advantage over the bare potential, $\tilde{V}_D$, for certain values of $R$. However, it is clear that there is quite complex structure in the variation of  $\tilde{V}_D^R$ with $R$ and there is no single value of $R$ for which all matrix elements obtain the smallest relative error.

 \begin{table}[htbp]
    \centering 
    \begin{tabular}{ cc } % Column formatting, @{} suppresses leading/trailing space
     \hline\hline
 $Z^{200}_{000}\quad$ & $\frac{1}{15\sqrt{\pi}}=0.037\,612\,\cdots$ \\
 $Z^{202}_{000}\quad$ & $\frac{1}{42\sqrt{2\pi}}=0.009\,498\,\cdots$\\
 $Z^{800}_{600}\quad$ & $-0.003\,908\,\cdots$ \\
 $Z^{644}_{444}\quad$ & $0.000\,462\,291\,\cdots$ \\
   \hline
         \hline\hline
    \end{tabular}
    \vspace*{3mm}
    \caption{Values of pure dipole matrix elements considered (see text)}
    \label{tab:ZRresults}
 \end{table}

To interpret these results it is useful to qualitatively classify the matrix elements into two categories: \\
\noindent \textit{Low order matrix elements}: These are matrix elements that involve  low order oscillator states, i.e.~those with quantum numbers much less than $M_x$ (i.e. the cases in Figs.~\ref{fig:R}(a) and (b)). For these cases the typical density variations are well-resolved on the quadrature grids and for both cases we see that  $R\approx\sqrt{2M_x}$ is the optimal value for obtaining such matrix elements with small relative error. This value appears to be universally good for low order matrix elements

\noindent \textit{High order matrix elements}: These are matrix elements that involve 
  oscillator states with quantum numbers comparable to $M_x$ (i.e.~the case in Figs.~\ref{fig:R}(c) and (d)). For these cases the typical density variations are rapid on the quadrature grids and $R\approx\sqrt{2M_x}$ is clearly not the optimal value for obtaining such matrix elements with small relative error.  The location of the minimum relative error (e.g.~$R\approx0.4\sqrt{2M_x}$ in Fig.~\ref{fig:R}(c), $R\approx0.65\sqrt{2M_x}$ in Fig.~\ref{fig:R}(d)) appears to vary appreciably with the particular high order matrix element, so that there is no universally good value.
  
In what follows we will take $R=\sqrt{2M_x}$. We make this choice because this appears to universally improve the accuracy of the low order matrix elements by at least several orders of magnitude over the uncorrected values, while only having a minor detrimental effect on the accuracy of the higher order matrix elements.  The cases presented in this section have been for the reference value ($N_k^0=2M_x$) of Fourier grid points. If additional Fourier points are added the best $R$ value for the low order matrix elements is $R\approx\sqrt{2N_k}$. This can be understood as follows: the use of the corrected interaction introduces an infrared cutoff in Fourier space at the wavevector scale $k_{\rm{cut}}\sim1/R$, which for the reference case $N_k=N_k^0$ is approximately equal to the  spacing between $k$ grid points near $k=0$. As we increase $N_k$ the $k$ grid resolution improves, i.e. smaller wavevectors are resolved and a longer  $R$ value is needed to represent the correspond longer wavelengths.

\subsection{Energy convergence for a Gaussian density}
Ronen \etal\ \cite{Ronen2006a} have checked the accuracy of their numerics by evaluating the dipolar energy functional \cite{note5}
\begin{equation}
I_{D}=\int\int d^3x\,d^3x'\,V_D(\x-\x')n(\x')n(\x),
\end{equation}
for the case of $D=1$ and the Gaussian density
\begin{equation}
n(\x)=\frac{1}{4\pi^{3/2}}e^{-(x^2+y^2)/4-z^2},\label{GaussDen}
\end{equation}
for which the exact result is 
\begin{equation}
I_D=0.038\,670\,861 \,\cdots.
\end{equation}
The results of Ronen \etal~are shown in Table \ref{tab:Eresults}, and clearly reveal the large improvement they obtained by using the corrected dipolar interaction.

It is not possible to directly compare our harmonic oscillator approach since we do not have independent control of the spatial extent and number of grid points. However, we can vary $M_x$ and check convergence \cite{note6}. For this case we use an isotropic harmonic oscillator potential, so that the density (\ref{GaussDen}) cannot be simply related to any finite superposition of eigenmodes of $H_0$. Thus, we explicitly construct $n(\x)$ on the quadrature grid before performing the normal transformations to make $\Phi(\x_{\mathbf{s}})$. To calculate the energy functional we then evaluate
\begin{equation}
I_D=\sum_{\mathbf{s}}w_{\mathbf{s}}e^{2|\x_{\mathbf{s}}|^2}\Phi(\x_{\mathbf{s}})n(\x_{\mathbf{s}}).
\end{equation}

The results shown in Table \ref{tab:Eresults} reveal qualitatively similar behavior to those observed in by Ronen \etal, i.e.~we see that the accuracy of the calculation improves gradually as the number of points increases, and a rather dramatic improvement in the accuracy if the corrected dipolar interaction is used.

 \begin{widetext}
 
 \begin{table}[htbp]
    \centering 
    \begin{tabular}{ ccccc } % Column formatting, @{} suppresses leading/trailing space
     \hline\hline
   Results of Ref.~\cite{Ronen2006a}: &\multicolumn{4}{c}{Relative error} \\
        & $R\!=\!8, N\!=\!32\quad$ & $R\!=\!8, N\!=\!64\quad$ & $R\!=\!16, N\!=\!64\quad$ & $R\!=\!16, N\!=\!128$ \\
        \hline
        Using $\tilde{V}_D(\k)\qquad$ & $2.7\times10^{-3}$ & $2.7\times10^{-3}$ & $8.6\times10^{-5}$ & $8.6\times10^{-5}$ \\
        Using $\tilde{V}_D^R(\k)\qquad$ & $-1.1\times10^{-5}$ & $-1.1\times10^{-5}$ & $1.8\times10^{-8}$ & $-4.4\times10^{-14}$ \\
        \\
       \hline     \hline
 
        Our results:  &  \multicolumn{3}{c}{Relative error}\\
        & $M_x=16$ & $M_x=32$ & $M_x=64$ \\
        \hline
        Using $\tilde{V}_D(\k)\qquad$ & $-1.7\times10^{-2}$ & $-3.1\times10^{-3}$ & $-5.5\times10^{-4} $ & $ $ \\
        Using $\tilde{V}_D^R(\k)\qquad$ & $-2.9\times10^{-3}$ & $-1.9\times10^{-5}$ & $8.3\times10^{-9} $ & $$ \\ \\
      \hline\hline
    \end{tabular}
    \vspace*{3mm}
    \caption{Relative error of the dipole interaction energy. We compare to the results of Ronen \etal\ \cite{Ronen2006a} using a 3D FFT method  on a cubic grid of extent $[-R,R]$ with $N$ points in each direction}
    \label{tab:Eresults}
 \end{table}
 
\end{widetext}

\subsection{Pure dipole matrix element convergence}\label{PDMEconv}
\begin{figure}[htbp] %  figure placement: here, top, bottom, or page
   \centering
   \includegraphics[width=3.2in]{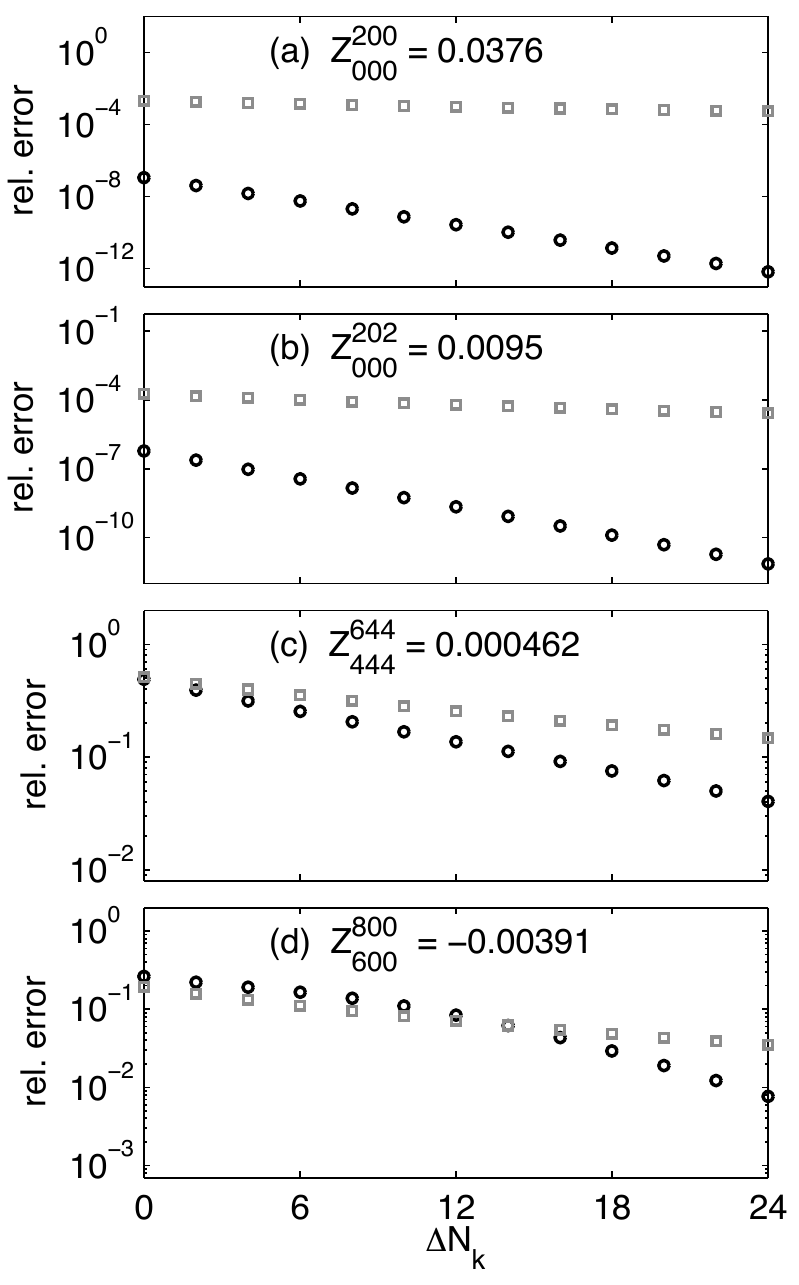} 
   \caption{Relative error in the pure dipole matrix elements (a) $Z^{200}_{000}$, (b)   $Z^{202}_{000}$, (c) $Z^{644}_{444}$and (d) $Z^{800}_{600}$, as $\Delta N_k$ is varied. Results: (grey squares) calculated using  $\tilde{V}_D(\k)$ ; (black circles) calculated using  $\tilde{V}_D^R(\k)$. Other parameters: $M_x=16$ and $R=\sqrt{2 N_k}$ (see text). }
   \label{fig:C}
\end{figure}

In this section we investigate the effect of increasing the number of $k$ grid points on the accuracy of pure dipolar matrix elements. Typical results for the relative error are shown in Fig.~\ref{fig:C}, with the corresponding exact matrix element values given in Table \ref{tab:ZRresults}. Figures~\ref{fig:C}(a) and (b) show the characteristic behavior for the low order matrix elements, indicating the general trend that these matrix elements improve considerably with $\Delta N_k$. For the higher order matrix elements [see Figs.~\ref{fig:C}(c) and (d)], the improvement in the relative error is much more gradual, but quite significant considering the rather low relative accuracy of these matrix elements in the reference configuration. The case seen in Fig.~\ref{fig:C}(d) shows that by increasing $\Delta N_k$ we can make the error in the corrected interaction matrix element smaller than the uncorrected value.

\subsection{Random state convergence}\label{rands}
\begin{figure}[htbp] %  figure placement: here, top, bottom, or page
   \centering
   \includegraphics[width=3.2in]{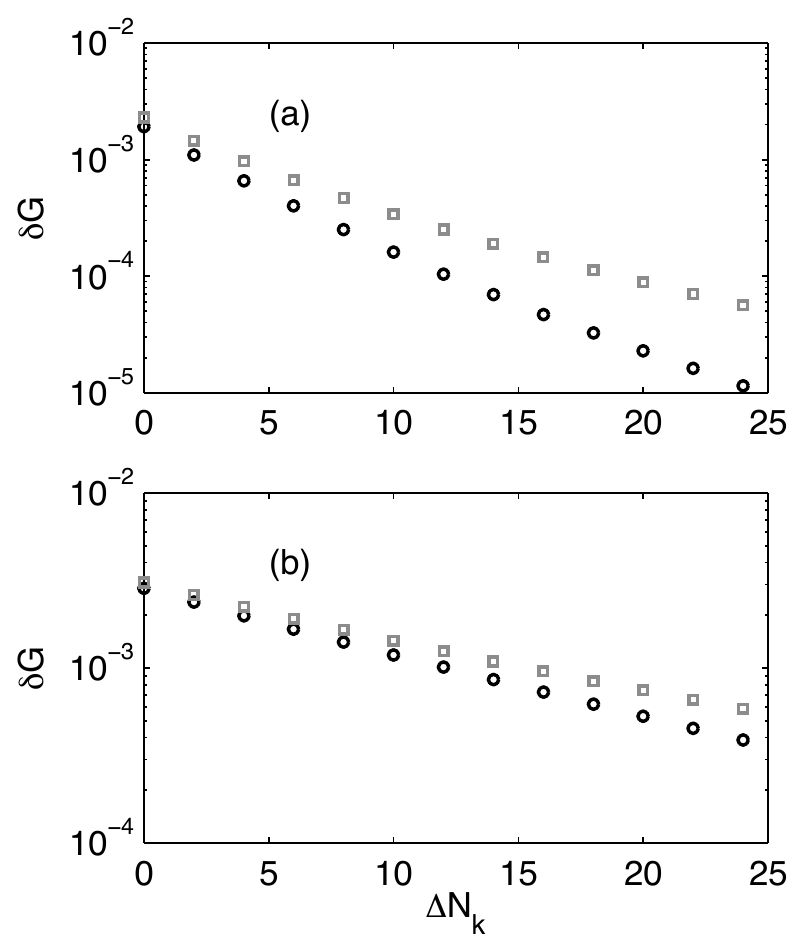} 
   \caption{Relative error in the random state matrix elements. Results: (grey square) calculated using  $\tilde{V}_D(\k)$; (black circles) calculated using  $\tilde{V}_D^R(\k)$. Results for a randomized state with (a) $M_x=10$ and (b) $M_x=30$.  }
   \label{fig:G}
\end{figure}

The pure dipole matrix elements considered so far are useful for understanding the general effects of using $\tilde{V}_D^R(\k)$ and changing $\Delta N_k$. However, for the purposes of understanding the PGPE in operation, a more appropriate test is to determine the nonlinear matrix elements, $G_{\bm{\sigma}}$, for a randomized state $c_{\bm{\sigma}}$. We can then determine the combined effect of altering $\tilde{V}_D^R(\k)$ and $\Delta N_k$ by examining 
%how close the complete set of matrix elements, $G_{\bm{\sigma}}^{(j)}$, are to the exact values, $G_{\bm{\sigma}}^{\rm{e}}$, according to the measure 
\begin{equation}
\delta G \equiv \frac{|\!|G_{\bm{\sigma}}-G_{\bm{\sigma}}^{A}|\!|^2}{|\!|G_{\bm{\sigma}}^{A}|\!|^2},\label{dG}
\end{equation}
where   $G_{\bm{\sigma}}$ refers to the approximate matrix elements,  $G_{\bm{\sigma}}^{A}$ refers to the more accurately calculated matrix elements (see below), and $|\!|\Lambda_{\bm{\sigma}}\,|\!|^2\equiv \sum_{\bm{\sigma}}|\Lambda_{\bm{\sigma}}|^2$. %The quantity $\delta G$ is relevant because 
The matrix elements $G_{\bm{\sigma}}$ determine the transitions between the bare spectral states in the PGPE (since $H_0$ is diagonal in that basis) and thus $\delta G$ measures the extent to which our approximate evaluation of $G_{\bm{\sigma}}$ matches the more accurate value $G_{\bm{\sigma}}^{A}$. We note that this differs from the earlier consideration of pure matrix elements because a large relative error in a small matrix element (typically the case for high order modes) has little effect on $\delta G$.

In Fig.~\ref{fig:G} we show results for $\delta G$ for cases where $G_{\bm{\sigma}}$ is evaluated using the bare and corrected dipole interaction, and for various values of $\Delta N_k$. Our pseudo-random state is reproducible, with procedure outlined in  Appendix \ref{RandStateA}. The accurate values, $G_{\bm{\sigma}}^{A}$, are calculated using our algorithm with $N_k=128$ quadrature points and $\tilde{V}_D^R$.

The results in Figs.~\ref{fig:G}(a) and (b) are for $M_x=10$ and $M_x=30$, respectively. In both cases the corrected dipole matrix element is more accurate, and converges more rapidly with $\Delta N_k$. For larger $M_x$ the convergence rate is less rapid, due to the increase in higher order matrix elements which our previous results show to converge more slowly.

\subsection{Propagation convergence}\label{PROPconv}

\begin{widetext}

 \begin{table}[htbp]
   \centering 
\begin{tabular}{c  c  |c | c  c c c c }
\hline\hline
Relative Tolerance & $\qquad\Delta N_k\qquad$ & Number of steps  & $\qquad\delta N\qquad$ & $\qquad\delta E\qquad$ & %$\delta E'$& 
$\qquad\delta L_z\qquad$ & $\qquad\delta X\qquad$  & $\qquad\delta X'\qquad$\\ 
\hline
 \hline 
10$^{-4}$ & 0 & 362 & -2.8$\times10^{-4}$ & 2.4$\times10^{-3}$ & 4.8$\times10^{-2}$ &   1.4$\times10^{-3}$ &   8.3$\times10^{-2}$ \\
 & 10  & 346 & -2.9$\times10^{-4}$ & 2.5$\times10^{-3}$ & 3.2$\times10^{-2}$ &   1.6$\times10^{-3}$ &   2.2$\times10^{-2}$ \\
 & 20  & 342 & -3.0$\times10^{-4}$ & 2.6$\times10^{-3}$ & 1.9$\times10^{-2}$ &   1.7$\times10^{-3}$ &   7.0$\times10^{-3}$ \\
 & 30  & 345 & -2.9$\times10^{-4}$ & 2.5$\times10^{-3}$ & 1.0$\times10^{-2}$ &   1.6$\times10^{-3}$ &   2.8$\times10^{-3}$ \\
 & 40 & 348 & -2.9$\times10^{-4}$ & 2.5$\times10^{-3}$ & 5.2$\times10^{-3}$ &   1.6$\times10^{-3}$ &   1.8$\times10^{-3}$ \\
 \hline
10$^{-5}$ & 0 & 570 & -2.8$\times10^{-5}$ & 2.5$\times10^{-4}$ & 4.8$\times10^{-2}$ &   1.5$\times10^{-5}$ &   8.0$\times10^{-2}$ \\
 & 10 & 551 & -2.9$\times10^{-5}$ & 2.5$\times10^{-4}$ & 3.2$\times10^{-2}$ &   1.5$\times10^{-5}$ &   2.1$\times10^{-2}$ \\
 & 20 & 554 & -2.9$\times10^{-5}$ & 2.5$\times10^{-4}$ & 1.8$\times10^{-2}$ &   1.5$\times10^{-5}$ &   5.2$\times10^{-3}$ \\
 & 30 & 554 & -2.9$\times10^{-5}$ & 2.5$\times10^{-4}$ & 1.0$\times10^{-2}$ &   1.5$\times10^{-5}$ &   1.1$\times10^{-3}$ \\
 & 40 & 557 & -2.9$\times10^{-5}$ & 2.5$\times10^{-4}$ & 5.0$\times10^{-3}$ &   1.5$\times10^{-5}$ &   1.6$\times10^{-4}$ \\
 \hline
10$^{-6}$ & 0 & 857 & -2.9$\times10^{-6}$ & 2.5$\times10^{-5}$ & 4.8$\times10^{-2}$ &   1.5$\times10^{-7}$ &   8.0$\times10^{-2}$ \\
 & 10 & 868 & -2.9$\times10^{-6}$ & 2.5$\times10^{-5}$ & 3.2$\times10^{-2}$ &   1.5$\times10^{-7}$ &   2.0$\times10^{-2}$ \\
 & 20 & 866 & -2.9$\times10^{-6}$ & 2.5$\times10^{-5}$ & 1.8$\times10^{-2}$ &   1.5$\times10^{-7}$ &   5.1$\times10^{-3}$ \\
 & 30 & 868 & -2.9$\times10^{-6}$ & 2.5$\times10^{-5}$ & 1.0$\times10^{-2}$ &   1.5$\times10^{-7}$ &   1.1$\times10^{-3}$ \\
 & 40 & 872 & -2.9$\times10^{-6}$ & 2.5$\times10^{-5}$ & 5.0$\times10^{-3}$ &   1.5$\times10^{-7}$ &   1.4$\times10^{-4}$ \\
 \hline
10$^{-7}$ & 0 & 1336 & -2.9$\times10^{-7}$ & 2.5$\times10^{-6}$ & 4.8$\times10^{-2}$ &   1.5$\times10^{-9}$ &   8.0$\times10^{-2}$ \\
& 10 & 1328 & -2.9$\times10^{-7}$ & 2.6$\times10^{-6}$ & 3.2$\times10^{-2}$ &   1.6$\times10^{-9}$ &   2.0$\times10^{-2}$ \\
 & 20 & 1344 & -2.9$\times10^{-7}$ & 2.6$\times10^{-6}$ & 1.8$\times10^{-2}$ &   1.6$\times10^{-9}$ &   5.1$\times10^{-3}$ \\
 & 30 & 1332 & -2.9$\times10^{-7}$ & 2.6$\times10^{-6}$ & 1.0$\times10^{-2}$ &   1.6$\times10^{-9}$ &   1.1$\times10^{-3}$ \\
 & 40 & 1341 & -2.9$\times10^{-7}$ & 2.5$\times10^{-6}$ & 5.0$\times10^{-3}$ &   1.6$\times10^{-9}$ &   1.4$\times10^{-4}$ \\
 \hline
10$^{-8}$ & 0 & 2079 & -2.9$\times10^{-8}$ & 2.5$\times10^{-7}$ & 4.8$\times10^{-2}$ &   1.5$\times10^{-11}$ &   8.0$\times10^{-2}$ \\
 & 10  & 2089 & -2.9$\times10^{-8}$ & 2.6$\times10^{-7}$ & 3.2$\times10^{-2}$ &   1.5$\times10^{-11}$ &   2.0$\times10^{-2}$ \\
 & 20  & 2085 & -2.9$\times10^{-8}$ & 2.6$\times10^{-7}$ & 1.8$\times10^{-2}$ &   1.5$\times10^{-11}$ &   5.1$\times10^{-3}$ \\
 & 30  & 2088 & -2.9$\times10^{-8}$ & 2.6$\times10^{-7}$ & 1.0$\times10^{-2}$ &   1.5$\times10^{-11}$ &   1.1$\times10^{-3}$ \\
 & 40  & 2092 & -2.9$\times10^{-8}$ & 2.6$\times10^{-7}$ & 5.0$\times10^{-3}$ &   1.5$\times10^{-11}$ &   1.4$\times10^{-4}$ \\
 \hline\hline
\end{tabular}
\caption{\label{convtable} 
Convergence properties of evolution algorithm. The relative error tolerance of the adaptive step Runge-Kutta algorithm, number of steps needed to obtain that error tolerance, and the quantitative measures $\delta N$, $\delta E$, $\delta L_z$ and $\delta X$ are shown (see text).   Other parameters: $T=1$, $C=500$, $D=500$, ${\epsilon}_{\rm{cut}}=23$ and the initial state is a thermalized state with energy ${E}=10.0$ (see text). All results computed using the corrected dipole interaction. 
}
\end{table}

\end{widetext}

Here we present some evolution convergence results for our algorithm.  
We have used an adaptive step Runge-Kutta-Fehlberg algorithm to evolve the dipolar PGPE with a specified relative error tolerance.
For all the results presented in the remainder of this paper we use the corrected dipole interaction so as to benefit from its generally more accurate evaluation of the matrix elements. 
Since computing the matrix elements for our harmonically trapped algorithm is of computational cost $O(M^{4/3})$ the development of higher order or more efficient propagation algorithms would be desirable (e.g.~see Refs.~\cite{Muruganandam2003,Adhikari2002,Xu2006,Chin2007}), although we do not address this issue further here.

We test our algorithm by propagating an initial state forward in time by an amount $T=1$. The system we consider has interaction parameters $C=500$ and $D=500$, and is in an isotropic trap potential with energy cut off  $\ecut=23$, for which $M=2024$ modes lie in the c-field region. To provide a useful analysis of the regime that the PGPE approach is normally used, we take an initial state of energy $E=10.0$ (as given by Eq.~({\ref{eq:EPGPE})) after it has been propagated to thermalize for 25 trap periods. This state has the desirable feature that all the modes of the field are appreciably occupied, and thus provides a more stringent test of the evolution.

In Table \ref{convtable} we examine the evolution convergence as we vary both the integration tolerance and $\Delta N_k$,  using the following measures:
\begin{eqnarray}
\delta N &=& 1- \sum_{j=1}^{M}|c_j(T)|^2,\label{dN}\\
\delta E &=& \frac{E[\cf(\x,T)]-E[\cf(\x,0)]}{E[\cf(\x,0)]},\label{dE}\\
\delta L_z &=& \frac{\langle L_z(T)\rangle-\langle L_z(0)\rangle }{\langle L_z(0)\rangle},\label{dL}\\
\delta X &=& \sum_{j=1}^{M}|c_j(T)-c^A_j(T)|^2,\label{dX}
\end{eqnarray}
i.e.~the change in normalization ($\delta N$), the relative change in energy ($\delta E$),  the relative change in the $z$ component of angular momentum ($\delta L_z$), and a difference measure of the final states ($\delta X$), where $c^A_j({T})$ are the mode amplitudes at time ${T}$ of a more accurate simulation (discussed below). The quantity  $\delta X$ provides a direct test of the field convergence at the final time. However, the other quantities considered relate to constants of motion, which are useful in practice as they provide a characterization of the accuracy without the need for running additional simulations.

\paragraph{Normalization} The dipolar PGPE formally preserves the normalization of the field. Our results in Table \ref{convtable} show that this quantity, as defined in Eq.~(\ref{dN}), is dependent on the tolerance of evolution algorithm, and is insensitive to the matrix element accuracy (i.e.~$\Delta N_k$).

\paragraph{Energy} The field energy is evaluated according to energy functional Eq.~(\ref{eq:EPGPE}). Unlike normalization, which can be calculated to numerical precision, the energy is limited to the precision with which we can evaluate the dipole energy. For the results in Table \ref{convtable} the energy functional is evaluated for the same value of $\Delta N_k$ as was used for the evolution under consideration.  These results reveal a similar convergence behavior to that observed for $\delta N$.
%
% Our results for $\delta E$ shown in Table \ref{convtable} are consistent with the accuracy of matrix elements in Secs.~\ref{PDMEconv} and \ref{rands}, i.e.~at best we might expect the relative energy to be conserved at the $10^{-4}$ level. 

\paragraph{Angular momentum}  For the dipolar system the anisotropic nature of the long-range interaction leads to interesting dynamics of the angular momentum, which we discuss further in Sec.~\ref{secL}. However, for the case of a spherical trap the $z$ component of angular momentum is conserved. To characterize this we evaluate
\begin{equation}
\langle L_z(t)\rangle = \int d^3x\,\cf^*(\x,t){L}_z\cf(\x,t),
\end{equation}
 where ${L}_z$ is the $z$ component of ${\mathbf{L}}=-i\hbar\x\times\bm{\nabla}$. Like normalization (and in contrast to the energy), the angular momentum can be evaluated efficiently and to numerical precision using the step operator formalism, as discussed in Ref.~\cite{Blakie2008a}. The results in Table \ref{convtable} show that $\delta L_z$ appears to converge quite slowly in $\Delta N_k$, and is conserved at the $10^{-3}$ level for our $\Delta N_k=40$ simulations. This may indicate an important consideration for the dipolar PGPE, and we discuss this further below.
 
% 
%  to $5.0\times10^{-3}$. We believe that this arising from residual projector effects which alter the Ehrenfest evolution from the (unprojected) continuum case \cite{Bradley}.
 
\paragraph{Field convergence} The quantity $\delta X$ indicates the extent to which the field evolution has converged. The results for $\delta X$ in Table \ref{convtable} have been computed by comparing each case to a more accurate calculation with  a relative tolerance of $10^{-9}$ and the same $\Delta N_k$ value. These results are insensitive to $\Delta N_k$ and show rapid convergence as the evolution tolerance is decreased. 
However, an important dependence on $\Delta N_k$ is revealed by computing $\delta X'$, defined as in Eq.~(\ref{dX}), but by comparing the against a $c^A(T)$ for a different (i.e.~larger) $\Delta N_k$ value. These results, presented in  Table \ref{convtable} for the case where the accurate solution uses a relative tolerance of $10^{-9}$ and $\Delta N_k=50$, reveal a much slower convergence in the parameter $\Delta N_k$, with a very weak dependence on evolution tolerance. This appears to be due to the rather slow convergence of the high energy matrix elements with $\Delta N_k$ as noted earlier (e.g. see Sec.~\ref{PDMEconv}). 
These results serve to illustrate an important point: Our algorithm in a fixed $\Delta N_k$ subspace is well-defined and displays good convergence primarily dependent on the evolution tolerance.

We note that the individual simulations reported in Table \ref{convtable} took between 3 minutes ($\sim350$ steps with $\Delta N_k=0$) and 2 hours ($\sim2000$ steps with $\Delta N_k=40$) using unoptimized single CPU code running on a shared cluster of 2.66GHz Clovertown Xeons.

%To realize a scheme that converges rapidly to the true solution would require a better quadrature for dealing with the dipole matrix elements, however as we show the next section for finite temperature applications even our le

%While we have strived to ensure highly accurate time propagation, statistical mechanical arguments suggest that this may be unnecessary and some additional \emph{noise} from imprecise evolution may assist the system explore the ensemble of available states more rapidly, as long as this noise does not effect the constants of motion describing the system. For our evolution algorithm the normalization of the field is not conserved and anecdotal evidence suggests that monitoring changes in normalization can be used as a summative assessment of the evolution accuracy, e.g.~for the simulation described above 
%the final state normalization was 0.99989.

\subsection{Convergence of thermodynamic predictions}
% The study we have presented thus far gives a rather broad characterization of the accuracy of our scheme for calculating dipole matrix elements, and analysis of the time evolution convergence. 
 
The important question we have yet to address is: what accuracy is required to perform a useful PGPE simulation? In general the answer to this question will depend on the particular application of interest, and in this final part  of the paper we will present some illustrative examples.
 
For deterministic applications, such as solving a $T=0$ Gross-Pitaevskii equation from a well-defined initial state, the small errors in the matrix elements will cause errors to accumulate leading to a practical time limit for the duration over which a calculation can be considered to be reliable. 
In contrast, the PGPE theory is typically operated in an ergodic regime of evolution, in which we only aim to specify or measure macroscopic features of the field.
%
%The results in the last section showed that small errors in the matrix elements lead to slightly different trajectories from the same initial state for different $\Delta N_k$.
%Indeed, we can view each $\Delta N_k$ case as being a dynamical system with nonlocal interaction which approximates the dipolar interaction quite well.  
An approximate treatment of the dipole interaction (e.g. all matrix elements at the $10^{-3}$ level of accuracy or better) would seem to be more than adequate for such applications, as long as our approach does not break important symmetries of the system, e.g. allowing constants of motion to change appreciably with time so that the system  relaxes to the wrong equilibrium state.
 
 To investigate these issues we simulate the evolution of the dipolar PGPE in a finite temperature regime, and explore how changing $\Delta N_k$ affects its predictions. To do this we prepare a random state of energy $E=10.0$, for an isotropic harmonic trap with $C=500$, $D=500$ and $\ecut=23$. We use this state as the initial condition for 8 simulations which differ in $\Delta N_k$ from 0 to 28. In each case we propagate the dipolar PGPE, using the adaptive step Runge-Kutta algorithm with a tolerance of $10^{-7}$, for $T=80\pi$ (i.e. 40 trap periods), saving the field at 1600 equally spaced times during the evolution.

\subsubsection{System width}
The randomly generated initial state used in the PGPE is an atypical (far from equilibrium state) and will evolve for some initial period until the system explores more typical microstates (i.e. rethermalizes).  After this initial period we can compute ensemble averages of equilibrium parameters by making use of the system's ergodicity.

A simple macroscopic parameter to compute is the mean system width, as characterized by the position variance,  e.g. $W_x(t)=\langle{x^2(t)}\rangle-\langle{x(t)}\rangle^2$ in the $x$ direction, where 
\begin{equation}
\langle x^n(t)\rangle = \int d^3x\,x^n|\cf(\x,t)|^2,
\end{equation}
is the instantaneous moment.
To make equilibrium predictions it is useful to calculate the  averaged width, which we  calculate using time-averaging, i.e.
the time averaged moment is given by
\begin{equation}
\overline{W}_x = \frac{1}{N_s}\sum_{j=1}^{N_s}W_x(t_j),
\end{equation}
where $N_s$ is the number of samples used.
 We avoid writing the similar expressions for $\overline{W}_y$ and $\overline{W}_z$. 

In what follows we let the system thermalize for the first $~10$ trap periods (in practice most large scale motion damps in the first few trap periods), and then perform time averaging using $N_s\sim1300$ states over the subsequent 32 trap period evolution.

 \begin{figure}[htbp] %  figure placement: here, top, bottom, or page
   \centering
   \includegraphics[width=3.2in]{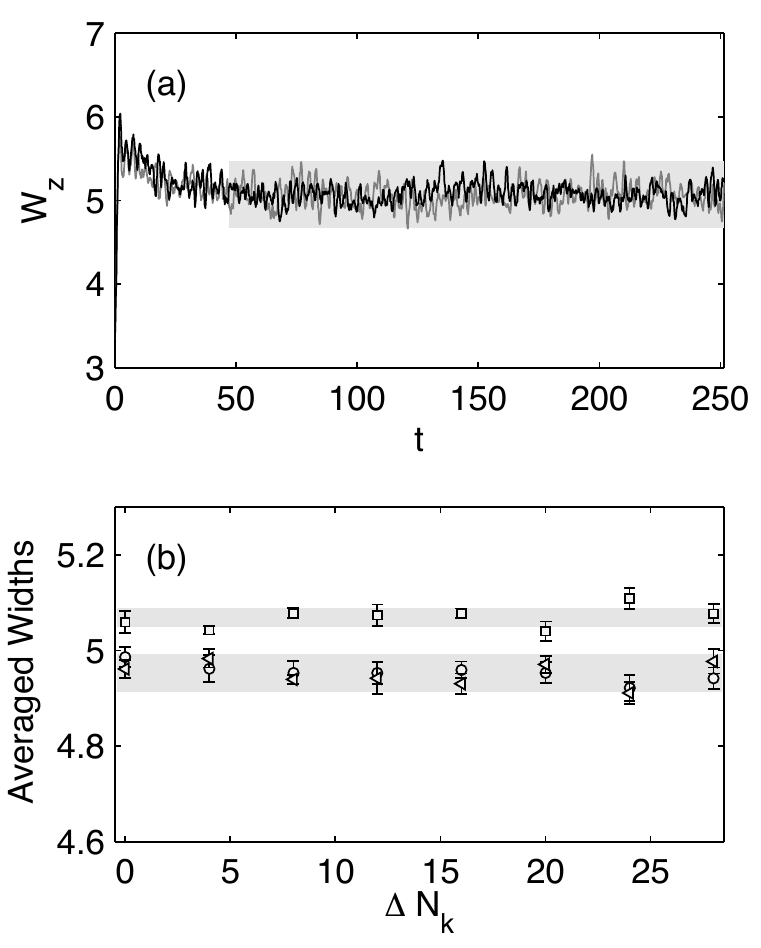} 
   \caption{(a) Evolution of $z$ width for simulation with $\Delta N_k=0$ (grey line) and $\Delta N_k=28$ (black line). (b) Dependence of the time averaged predictions for the variances of a dipolar Bose gas on $\Delta N_k$. Results for $\overline{W}_x$ (circles), $\overline{W}_y$ (triangles), and $\overline{W}_z$ (squares) in units of the harmonic oscillator length squared ($x_0^2=\hbar/m\omega$). Shaded region in (a) indicates the states used for time averaging. The shaded regions in (b) characterize the spread in results. All results computed using the corrected dipole interaction. }
   \label{fig:ThermConv}
\end{figure}

The results for the position width are shown in Fig.~\ref{fig:ThermConv}. Interestingly the width of the system in the $z$ direction is greater than the $x$ and $y$ directions even though the system is in an isotropic harmonic trap. This asymmetry arises from the polarization of the dipoles in the $z$ direction  which causes the system to slightly elongate to reduce the dipolar interaction energy. 
We note that there is no clear change in the results with $\Delta N_k$ \cite{note7} and the improved accuracy associated with increasing $\Delta N_k$ is clearly unimportant in this case. 
The states over which time averaging is performed are indicated by the shaded region in Fig.~\ref{fig:ThermConv}(a). Interestingly the breadth of this region (chosen to match the range of the equilibrium width dynamics) is 20 times larger than the shaded region shown in Fig.~\ref{fig:ThermConv}(b) to indicate the spread in the averaged $z$ width results. This suggests that while the width dynamics are quite appreciable, the averages are very well-defined. Longer time averages could be used to further refine these predictions.
We also note that the larger variation in the $x$ and $y$ variances seem to  result from strong collective dynamics associated with the non-conservation of angular momentum, which we discuss below.

\subsubsection{Angular momentum evolution}\label{secL}
 The anisotropic (non-central) nature of the dipole interaction means that angular momentum is not conserved even for the case of a spherical external potential. 
Indeed, as can be shown (see Appendix \ref{Ehrenfest}) the evolution of the angular momentum is given by
\eqn{
\frac{d\langle L_x\rangle}{dt}&=&-4\pi D\int d^3k\,\tilde{n}(\kk)\tilde{n}(-\kk)\frac{k_yk_z}{\mathbf{|k|}^2},\label{Lx1}\\
\frac{d\langle L_y\rangle}{dt}&=&4\pi D\int d^3k\,\tilde{n}(\kk)\tilde{n}(-\kk)\frac{k_zk_x}{\mathbf{|k|}^2},\label{Ly1}\\
\frac{d\langle L_z\rangle}{dt}&=&0,\label{Lz1}
}
(for the isotropic trap case), revealing that the invariance of rotations about the polarization direction leads to conservation of the $z$ component of $\mathbf{L}$.
This motivated the definition of  $\delta L_z$ as a numerical check in Sec.~\ref{PROPconv}.

  \begin{figure}[htbp] %  figure placement: here, top, bottom, or page
   \centering
   \includegraphics[width=3.2in]{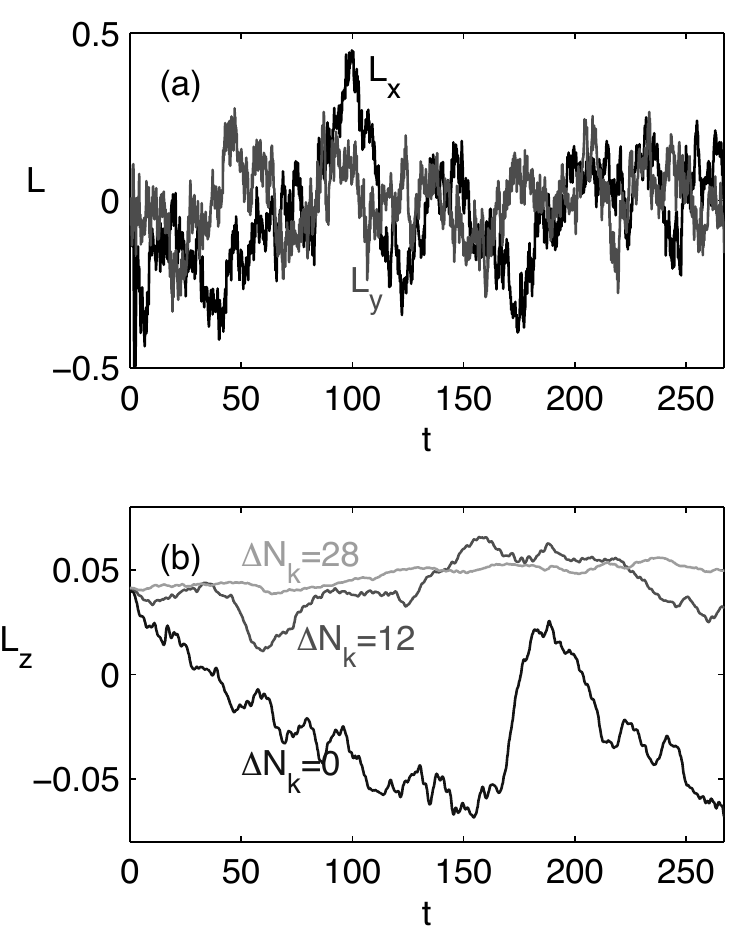} 
   \caption{Angular momentum evolution. (a) $x$ and $y$  components of angular momentum in units of $\hbar$ over a small time segment of the simulation for  $\Delta N_k=0$. (b) The $z$ component of angular momentum of the whole evolution for simulations of various $\Delta N_k$ values. Parameters are the same as for the results presented in Fig.~\ref{fig:ThermConv} and time is measured in units of inverse trap frequency ($1/\omega$). All results computed using the corrected dipole interaction. }
   \label{fig:Lbehav}
\end{figure}
 
 In Fig.~\ref{fig:Lbehav}(a) we show the evolution of the $x$ and $y$ components of angular momentum for our dipolar simulations. As suggested by Eqs.~(\ref{Lx1})-(\ref{Lz1}), the $x$ and $y$ components of angular momentum show strong dynamics. These dynamics are a contributing factor to the slightly larger spread in results for the position variance in the $x$ and $y$ directions relative to the $z$ direction, as seen in Fig.~\ref{fig:ThermConv}(b). 
 
In Fig.~\ref{fig:Lbehav}(b)  we examine the evolution of $L_z$. According to Eq.~(\ref{Lz1}) $L_z$ should be conserved, and so the dynamics of this quantity indicates inaccuracy in our algorithm.
 The various curves in Fig.~\ref{fig:Lbehav}(b) indicate that as $\Delta N_k$ increases, the drift in $L_z$ decreases. In some applications of the dipolar PGPE theory, e.g.~ in studies of vortices, careful attention to $L_z$ conservation will be prudent and will demand the use of a large $\Delta N_k$. However, for many  applications the quasi-stationary behavior of $L_z$ observed in the $\Delta N_k=0$ case will be adequate to make reliable predictions (e.g. our position variance results appear insensitive to $\Delta N_k$).

\section{Conclusions}
In this paper we have presented a numerical method that allows us to extend the PGPE theory to include long-range dipolar interactions.  We have used a range of tests to characterize the numerical accuracy of our scheme and the convergence with increasing order of $k$-space quadrature grid. These results show that use of the corrected dipole potential is a significant improvement,  and that our approach is sufficiently accurate to make reliable physical predictions in the context of finite temperature c-field calculations. 
Many aspects of the formalism we have developed are quite general and would easily allow us to apply the method to a wider class of long-range interactions.

\subsection*{Acknowledgments}
PBB is supported by NZ-FRST contract NERF-UOOX0703, ASB is supported by NZ-FRST contract  UOOX0801. CT and MJD acknowledge the financial support of the Australian Research Council Centre of Excellence for Quantum-Atom Optics.  PBB would like to thank Dr Shai Ronen for useful feedback on the manuscript.
 
\appendix
\section{Randomized state}\label{RandStateA}
We generate a pseudo random state based on 
a \textit{linear congruential generator}, with recurrence relation 
\begin{equation}
X_{n+1}=\mod_m(aX_{n}+c),\label{LCG}
\end{equation}
with $a=16807$, $c=0$, and $m=2^{31}-1$.

We prepare a set of complex random numbers defining the classical field, $c_{\alpha\beta\gamma}$. To do this we map the quantum number tuples $\{\alpha,\beta,\gamma\}$ to a unique integer value, $n$, according to
\begin{equation}
n = \alpha+M_x\beta+M_x^2\gamma.
\end{equation}
We then specify our classical field state as
\begin{equation}
c_n = \frac{1}{m}\left(X^{(1)}_n+iX^{(2)}_n\right),
\end{equation}
where $X^{(1)}_n$ is the sequence generated by (\ref{LCG}) with seed $X^{(1)}_0=10^8$, and $X^{(2)}_n$ is the sequence generated by (\ref{LCG}) with $X^{(2)}_0=10^9$. Thus, we have 
\begin{eqnarray}
c_0\equiv c_{000}&=&0.0466 + 0.4657i, \\
  c_1\equiv c_{100}&=&0.6369 + 0.3693i,\\
  c_2\equiv c_{200}&=&  0.8143 + 0.1432i,\\
  \cdots
  \end{eqnarray}

\section{Angular momentum Ehrenfest relation}\label{Ehrenfest}
Given angular momentum operator $\mathbf{L}=-i\hbar\mathbf{x}\times\nabla$,
the standard Ehrenfest result for the GPE angular momentum is given by
\eq{e1}{
i\hbar\frac{d\langle \mathbf{L}\rangle}{dt}=\langle \mathbf{L}[V_0+\Phi]\rangle.
}
The effect of the harmonic trap potential, $V_0$, on the angular momentum evolution is well-understood. Here we will focus on the case of an isotropic trap (i.e.~$[\mathbf{L},V_0]=\mathbf{0}$) so that  the evolution arises from the effective dipole potential, $\Phi$, i.e.  $i\hbar {d\langle \mathbf{L}\rangle}/{dt}=\langle \mathbf{L}\Phi\rangle$.
 
 Taking the Fourier transformed form of $\Phi$ (see Eq.~(\ref{Phi2}))
 \eq{LV}{
\langle \mathbf{L}\Phi\rangle=\int d^3x\,\cf^*(\x)\int d^3k\,\tilde{V}_D(\kk)\tilde{n}(\kk)\left(\mathbf{L}e^{i\kk\cdot\x}\right)\cf(\x).
}
We can use the self-duality of angular momentum operators under Fourier transform, that is
\eq{Lk}{
\int d^3k\,\mathbf{L}e^{i\kk\cdot\x}=-\int d^3k\,\tilde{\mathbf{L}} e^{i\kk\cdot\x},
}
where $\tilde{\mathbf{L}}$ is the representation of angular momentum in $k$-space, i.e. $\tilde{L}_{k_z}=-i\hbar(k_x\partial_{k_y}-k_y\partial_{k_x})$. 
We then find
\eqn{\label{Lk1}
\langle \mathbf{L}\Phi\rangle&=&\int d^3k\,\tilde{n}(-\kk)\tilde{\mathbf{L}}\left(\tilde{V}_D(\kk)\tilde{n}(\kk)\right), \\
%&=&\mint{\kk}\tilde{n}(-\kk)\tilde{n}(\kk)\tilde{\mathbf{L}}\tilde{V}_D(\kk)+\tilde{n}(-\kk)\tilde{V}_D(\kk)\tilde{\mathbf{L}}\tilde{n}(\kk)\nonumber\\
%&=&\mint{\kk}\tilde{n}(-\kk)\tilde{n}(\kk)\tilde{\mathbf{L}}\tilde{V}_D(\kk)+\frac{1}{2}\tilde{n}(-\kk)\tilde{V}_D(\kk)\tilde{\mathbf{L}}\tilde{n}(\kk)+\frac{1}{2}\tilde{n}(\kk)\tilde{V}_D(\kk)\tilde{\mathbf{L}}\tilde{n}(-\kk)\nonumber\\
&=&\int d^3k\,\tilde{n}(-\kk)\tilde{n}(\kk)\tilde{\mathbf{L}}\tilde{V}_D(\kk)\\
&&+\int d^3k\,\frac{1}{2}\tilde{V}_D(\kk)\tilde{\mathbf{L}}(\tilde{n}(-\kk)\tilde{n}(\kk)),\nonumber\\
&=&\frac{1}{2}\int d^3k\,\tilde{n}(-\kk)\tilde{n}(\kk)\tilde{\mathbf{L}}\tilde{V}_D(\kk),
}
so that
\eq{dLdt}{
i\hbar \frac{d\langle \mathbf{L}\rangle}{dt}=\int d^3k\,\frac{\tilde{n}(\kk)\tilde{n}(-\kk)}{2}\tilde{\mathbf{L}}\tilde{V}_D(\kk).
}
We now make use of the Cartesian components of $\tilde{\mathbf{L}}$ in spherical co-ordinates:
\eqn{
-i\tilde{L}_{k_x}/\hbar&=&\frac{\cos\phi_k}{\tan\theta_k}\frac{\partial}{\partial \phi_k}+\sin\phi_k\frac{\partial}{\partial \theta_k}\label{Lkx},\\
-i\tilde{L}_{k_y}/\hbar&=&\frac{\sin\phi_k}{\tan\theta_k}\frac{\partial}{\partial \phi_k}-\cos\phi_k\frac{\partial}{\partial \theta_k}\label{Lky},\\
\tilde{L}_{k_z}&=&-i\hbar\frac{\partial}{\partial \phi_k}\label{Lkz},
}
where  $\phi_k$ is the azimuthal angle from $k_x$ in the $k_x$--$k_y$ plane. For the dipolar potential, we find
\eqn{
\tilde{L}_{k_x}\tilde{V}_D(\kk)&=&-8\pi i\hbar D\frac{k_yk_z}{\mathbf{|k|}^2}\label{LkxVd},\\
\tilde{L}_{k_y}\tilde{V}_D(\kk)&=&8\pi i\hbar D\frac{k_zk_x}{\mathbf{|k|}^2}\label{LkyVd},\\
\tilde{L}_{k_z}\tilde{V}_D(\kk)&=&0\label{LkzVd},
}
and the angular momentum equations
\eqn{
\frac{d\langle L_x\rangle}{dt}&=&-4\pi D\int d^3k\,\tilde{n}(\kk)\tilde{n}(-\kk)\frac{k_yk_z}{\mathbf{|k|}^2},\label{Lx}\\
\frac{d\langle L_y\rangle}{dt}&=&4\pi D\int d^3k\,\tilde{n}(\kk)\tilde{n}(-\kk)\frac{k_zk_x}{\mathbf{|k|}^2},\label{Ly}\\
\frac{d\langle L_z\rangle}{dt}&=&0,\label{Lz}
}
which should provide useful consistency conditions for numerical simulations. $\partial \langle L_z\rangle/\partial t= 0$ is expected from the cylindrical symmetry of $V_D(\x)$ about the polarization axis.
We can also see from Eq.~(\ref{Lx}) that if $\tilde{n}(\kk)=\tilde{n}(k_x,|k_y|,|k_z|)$ i.e. is reflection symmetric in the $k_y$ and $k_z$ directions, then $d\langle L_x\rangle/dl t\equiv 0$. Similarly, if $\tilde{n}(\kk)=\tilde{n}(|k_x|,k_y,|k_z|)$, then $d\langle L_y\rangle/d t\equiv 0$. We note that $\tilde{n}(-\kk)=\tilde{n}(\kk)$ holds when $n(-\x)=n(\x)$ so that eigenstates of parity will conserve $\mathbf{L}$. Consequently the evolution of a spherically symmetric state into a cylindrically symmetric state should conserve angular momentum. We have not included boundary terms in this derivation which arise from the projector, and future work will be to assess at what level they may contribute (e.g. see \cite{Bradley})

\section{Analytic evaluation of the pure dipole matrix elements}\label{dipoleME}
In this appendix we derive an analytical expression for the pure dipolar matrix elements, as given by Eq.~(\ref{puredipmeZ}):
\begin{eqnarray}
Z^{\delta \epsilon \zeta}_{\alpha \beta \gamma}&=&C^{\delta \epsilon \zeta}_{\alpha \beta \gamma}\int d^3x\, d^3x'\, e^{-(x^2+y^2+z^2)} H^{\delta \epsilon \zeta}_{\alpha \beta \gamma}({\bf x}) \nonumber \\
& \times&V_D({\bf x}- {\bf x^{\prime}}) e^{-(x^{\prime 2}+y^{\prime 2}+z^{\prime 2})}H_{\alpha \beta \gamma}^{\alpha \beta \gamma}({\bf x^{\prime}}),
\end{eqnarray}
where 
\begin{eqnarray}
C^{\delta \epsilon \zeta}_{\alpha \beta \gamma}=h_{\delta} h_{\epsilon} h_{\zeta} h_{\alpha}^3 h_{\beta}^3 h_{\gamma}^3,
\end{eqnarray}
and
\begin{eqnarray}
H^{\delta \epsilon \zeta}_{\alpha \beta \gamma}({\bf x})=H_{\delta}(x) H_{\epsilon}(y) H_{\zeta}(z)H_{\alpha}(x) H_{\beta}(y) H_{\gamma}(z).
\end{eqnarray}
Using the convolution theorem with
%\begin{eqnarray}
%Z^{\delta \epsilon \zeta}_{\alpha \beta \gamma}&=&C^{\delta \epsilon \zeta}_{\alpha \beta \gamma}\int d{\bf x} e^{-(x^2+y^2+z^2)} H^{\delta \epsilon \zeta}_{\alpha \beta \gamma}({\bf x}) \nonumber \\
%&\times&{ \cal F}^{-1} \left\{ {\cal F} \left\{V_D({\bf x}- {\bf x^{\prime}})\right\} {\cal F} \left\{e^{-(x^{\prime 2}+y^{\prime 2}+z^{\prime 2})}H_{\alpha \beta \gamma}^{\alpha \beta \gamma}({\bf x^{\prime}})\right\}\right\} \nonumber \\
%\end{eqnarray}
%where
\begin{eqnarray}
{\cal F} \left\{V_D({\bf x}- {\bf x^{\prime}})\right\} =\frac{4\pi D}{3} \left(\frac{3k_z^2}{k_{x}^2+k_y^2+k_z^2}-1\right),
\end{eqnarray}
and
\begin{eqnarray}
 &{\cal F}& \left\{e^{-(x^{\prime 2}+y^{\prime 2}+z^{\prime 2})}H_{\alpha \beta \gamma}^{\alpha \beta \gamma}({\bf x^{\prime}})\right\}= \nonumber \\ & &{\tilde C}_{\alpha \beta \gamma}e^{-(k_{x}^2+k_y^2+k_z^2)/4}L_{\alpha}\left(\frac{k_x^2}{2}\right)L_{\beta}\left(\frac{k_y^2}{2}\right)L_{\gamma}\left(\frac{k_z^2}{2}\right),\nonumber \\
\end{eqnarray}
where
\begin{eqnarray}
{\tilde C}_{\alpha \beta \gamma}=\alpha ! \beta! \gamma! (-1)^{3(\alpha+\beta+\gamma)}(i)^{2(\alpha+\beta+\gamma)}(2)^{-\frac{3}{2}+\alpha+\beta+\gamma}.
\end{eqnarray}
the dipolar matrix elements can be evaluated from
\begin{eqnarray}
Z^{\delta \epsilon \zeta}_{\alpha \beta \gamma}&=&\frac{4 \pi D}{3}B^{\delta \epsilon \zeta}_{\alpha \beta \gamma}\int d^3k\, e^{-(k_x^2+k_y^2+k_z^2)/2} \left(\frac{3k_z^2}{k_{x}^2+k_y^2+k_z^2}-1\right) \nonumber \\
&\times&L_{\alpha}\left(\frac{k_x^2}{2}\right)L_{\beta}\left(\frac{k_y^2}{2}\right)L_{\gamma}\left(\frac{k_z^2}{2}\right)k_x^{|\alpha-\delta|}k_y^{|\beta-\epsilon|}k_z^{|\gamma-\zeta|} \nonumber \\
&\times& L_{{\rm Min}[\alpha,\delta]}^{|\alpha-\delta|}\left(\frac{k_x^2}{2}\right)L_{{\rm Min}[\beta,\epsilon]}^{|\beta-\epsilon|}\left(\frac{k_y^2}{2}\right)L_{{\rm Min}[\gamma,\zeta]}^{|\gamma-\zeta|}\left(\frac{k_z^2}{2}\right),
\label{Z_laguerre}
\end{eqnarray}
where
\begin{eqnarray}
B^{\delta \epsilon \zeta}_{\alpha \beta \gamma}&=&C^{\delta \epsilon \zeta}_{\alpha \beta \gamma}{\tilde C}_{\alpha \beta \gamma}(-1)^{{\rm Max}[\alpha,\delta]+{\rm Max}[\beta,\epsilon]+{\rm Max}[\gamma,\zeta]} \nonumber \\
&\times&i^{\alpha+\beta+\gamma+\delta+\epsilon+\zeta}2^{-\frac{3}{2}+{\rm Min}[\alpha,\delta]+{\rm Min}[\beta,\epsilon]+{\rm Min}[\gamma,\zeta]} \nonumber \\
&\times& {\rm Min}[\alpha,\delta]! {\rm Min}[\beta,\epsilon]!{\rm Min}[\gamma,\zeta]!.  
\end{eqnarray}
Expressing the associated Laguerre polynomials in terms of a finite sum: 
\begin{eqnarray}
L_n^k(x)=\sum_{m=0}^n(-1)^m\frac{(n+k)!}{(n-m)!(k+m)!m!}x^m,
\end{eqnarray}
we find  that if  $\alpha-\delta$ or $\beta-\epsilon$ or $\gamma-\zeta$ is odd then $Z^{\delta \epsilon \zeta}_{\alpha \beta \gamma}=0$. When $\alpha-\delta$ and $\beta-\epsilon$ and $\gamma-\zeta$ are even Eq. (\ref{Z_laguerre}) {\it reduces} to
\begin{widetext}
\begin{eqnarray}
Z^{\delta \epsilon \zeta}_{\alpha \beta \gamma}&=&-\frac{8 \pi^2 D}{3}{\tilde B}^{\delta \epsilon \zeta}_{\alpha \beta \gamma} \sum_{j_1=0}^{\alpha}\sum_{j_2=0}^{\beta}\sum_{j_3=0}^{\gamma}\sum_{j_4=0}^{{\rm Min}[\alpha,\delta]}\sum_{j_5=0}^{{\rm Min}[\beta,\epsilon]}\sum_{j_6=0}^{{\rm Min}[\gamma,\zeta]} \left[\prod_{j_7=0,2,4...}^{2j_{25}+|\beta-\epsilon|-2}\left(2(j_{12}+j_{45})+|\alpha-\delta|+|\beta-\epsilon|-j_7\right)\right]^{-1} \nonumber \\
&\times&
 \frac{(-1)^j 2^{\frac{1}{2}(1-2j_{14}+|\beta-\epsilon|+|\gamma-\zeta|)}\left(2j-6j_{36}+|\alpha-\delta|+|\beta-\epsilon|-2|\gamma-\zeta|\right)\left(-1+2j_{14}+|\alpha-\delta|\right)!!\left(-1+2j_{25}+|\beta-\epsilon|\right)!!} {\left(j_1!j_2!j_3!\right)^2j_4!j_5!j_6!(\alpha-j_1)!(\beta-j_2)!(\gamma-j_3)!(j_4+|\alpha-\delta|)!(j_5+|\beta-\epsilon|)!(j_6+|\gamma-\zeta|)!}\nonumber \\
 &\times&\frac{\Gamma\left[1+j-j_{36}+|\alpha-\delta|/2+|\beta-\epsilon|/2\right]\Gamma\left[j_{36}+(1+|\gamma-\zeta|)/2\right]}{(j_{14}+|\alpha-\delta|/2)!\left(3+2j+|\alpha-\delta|+|\beta-\epsilon|+|\gamma-\zeta|\right)({\rm Min}[\alpha,\delta]-j_4)!({\rm Min}[\beta,\epsilon]-j_5)!({\rm Min}[\gamma,\zeta]-j_6)!} 
 \label{Z_sum}
\end{eqnarray}
\end{widetext}
%\begin{widetext}
%\begin{eqnarray}
%Z^{\delta \epsilon \zeta}_{\alpha \beta \gamma}&=&\frac{4 \pi D}{3}{\tilde B}^{\delta \epsilon \zeta}_{\alpha \beta \gamma} \sum_{j_1=0}^{\alpha}\sum_{j_2=0}^{\beta}\sum_{j_3=0}^{\gamma}\sum_{j_4=0}^{{\rm Min}[\alpha,\delta]}\sum_{j_5=0}^{{\rm Min}[\beta,\epsilon]}\sum_{j_6=0}^{{\rm Min}[\gamma,\zeta]}  \nonumber \\
%&\times&
 %\frac{(-1)^{j}}{(\alpha-j_1)!(\beta-j_2)!(\gamma-j_3)!({\rm Min}[\alpha,\delta]-j_4)!({\rm Min}[\beta,\epsilon]-j_5)!({\rm Min}[\gamma,\zeta]-j_6)!} \nonumber \\
 %&\times&\frac{1}{(j_1!j_2!j_3!)^2j_4!j_5!j_6!(|\alpha-\delta|+j_4)!(|\beta-\epsilon|+j_5)!(|\gamma-\zeta|+j_6)!2^j} \nonumber \\
 %&\times& \int d{\bf k} e^{-(k_x^2+k_y^2+k_z^2)/2} \left(\frac{3k_z^2}{k_{x}^2+k_y^2+k_z^2}-1\right)
%k_x^{|\alpha-\delta|+2j_1+2j_4}k_y^{|\beta-\epsilon|+2j_2+2j_5}k_z^{|\gamma-\zeta|+2j_3+2j_6} 
%\end{eqnarray}
%\end{widetext}
where
\begin{eqnarray}
{\tilde B}^{\delta \epsilon \zeta}_{\alpha \beta \gamma}&=&B^{\delta \epsilon \zeta}_{\alpha \beta \gamma}\alpha!\beta!\gamma! ({\rm Min}[\alpha,\delta]+|\alpha-\delta|)! \nonumber \\
&\times& ({\rm Min}[\beta,\epsilon]+|\beta-\epsilon|)!({\rm Min}[\gamma,\zeta]+|\gamma-\zeta|)!, \nonumber \\
\end{eqnarray}
$j=j_1+j_2+j_3+j_4+j_5+j_6$ and $j_{ab}=j_a+j_b$.

Equation (\ref{Z_sum}) can be readily evaluated and serves a direct comparison for the numerical integration.

\bibliographystyle{apsrev}
\bibliography{dipolar}

\begin{thebibliography}{54}
\bibitem[{not({\natexlab{a}})}]{note1}
\bibinfo{howpublished}{{Other sets of modes defining $\rC$ are permissible and
  easily implemented in our formalism. Here we focus on the single particle
  energy cut off because of its simplicity and wide range of applicability.}}
\bibitem[{not({\natexlab{b}})}]{note2}
\bibinfo{howpublished}{{This is not a restriction of the method, and without
  this simplification we need to carry cumbersome notation indicating different
  mode functions, quadratures, and numbers of points for each dimension.}}
\bibitem[{not({\natexlab{c}})}]{note3}
\bibinfo{howpublished}{{Since a polynomial of degree $2N-1$ is integrated
  exactly using an $N$-point quadrature.}}
\bibitem[{not({\natexlab{d}})}]{note4}
\bibinfo{howpublished}{{Here we form $e^{2| {\mathbf{x}}_{ijk}|^2}| {\cf}|^{2}
  {\cf}$ as this corresponds to the polynomial ($P$) required for the
  quadrature (see \ref{Gabc}).}}
\bibitem[{not({\natexlab{e}})}]{note5}
\bibinfo{howpublished}{{Formally this energy functional should have a factor of
  $1/2$ in front of it, as is formally shown in Ronen \etal~. However, this is
  a typo as the results they give do not include this factor.}}
\bibitem[{not({\natexlab{f}})}]{note6}
\bibinfo{howpublished}{{As noted at the beginning of this section, convergence
  with $M_x$ is not useful for PGPE applications. However, we perform this
  comparison here to more generally characterize our spectral approach and to
  compare against established results in the literature.}}
\bibitem[{not({\natexlab{g}})}]{note7}
\bibinfo{howpublished}{{The error bars are assigned by calculating the
  correlation time of the signal, e.g. $W_x(t)$ for which we can estimate the
  number of statistically independent samples $N^*_s<N_s$, and we then assign
  an error under the assumption that this quantity is Gaussian distributed,
  which appears reasonable from histograms of the signal.}}
\end{thebibliography}

\end{document}